%% file: main-arXiv-Recall.tex
\documentclass[sigconf, 10pt]{acmart}

\usepackage{hyperref}

\setcopyright{none}
\renewcommand\footnotetextcopyrightpermission[1]{}
\usepackage{soul}

\usepackage{multirow}
\usepackage[utf8]{inputenc}
\usepackage[ruled, lined, longend, linesnumbered]{algorithm2e}
\usepackage{amsmath,multicol}
\usepackage{subcaption,textcomp,comment}
\usepackage{threeparttable}

\usepackage{booktabs}
\usepackage{multirow}

\usepackage{array}
\newcolumntype{L}[1]{>{\raggedright\let\newline\\\arraybackslash\hspace{0pt}}m{#1}}
\newcolumntype{C}[1]{>{\centering\let\newline\\\arraybackslash\hspace{0pt}}m{#1}}
\newcolumntype{R}[1]{>{\raggedleft\let\newline\\\arraybackslash\hspace{0pt}}m{#1}}

\newcommand{\mwx}[1]{\textbf{\color{red}{#1}}}

\newcommand{\service}[0]{\texttt{Recall}\xspace}
\newcommand{\sys}[0]{\texttt{Recall}\xspace}

\input{header-lin}

\begin{document}

\title[\sys: Empowering Multimodal Embedding for Edge Devices]{\sys: Empowering Multimodal Embedding \\for Edge Devices}

\author{Dongqi Cai, Shangguang Wang, Chen Peng, Zeling Zhang, Mengwei Xu*}

\input{sec-abstract}

\maketitle

\input{todo}
\input{sec-intro}

\input{sec-bkgnd.tex}
\input{sec-design.tex}

\input{sec-impl.tex}

\input{sec-eval.tex}

\input{sec-related.tex}

\input{sec-conclusion.tex}

\balance
\bibliographystyle{IEEEbib}
\bibliography{bib/ref-cdq}

\end{document}

%% file: header-lin.tex
\usepackage{wrapfig}
\usepackage{comment}
\usepackage{colortbl}

\usepackage{color,soul}
\usepackage{graphics}
\usepackage{hyperref}
\usepackage{lipsum}
\usepackage{tikz}
\usepackage{enumitem}	
\usepackage{listings} 

\usepackage{booktabs}	
\usepackage{lipsum}

\usepackage{capt-of}	

\usepackage{todonotes}  
\usepackage{subcaption} 

\definecolor{refkey}{rgb}{0,0,1}
\definecolor{labelkey}{rgb}{0,0,1}

\usepackage{cleveref}
\AtBeginDocument{\DeclareCaptionSubType{lstlisting}}
\crefname{sublstlisting}{listing}{listings}
\Crefname{sublstlisting}{Listing}{Listings}

\usepackage{mathrsfs}	
\usepackage{amsfonts}

\usepackage[export]{adjustbox} 

\usepackage{boldline}					

\usepackage{multirow}

\renewcommand{\paragraph}[1]{\vskip 3pt\noindent\textbf{#1 }}	 

  {\begin{list}{$\bullet$}%
     {\setlength{\parsep}{0pt}%
      \setlength{\topsep}{0pt}%
      \setlength{\itemsep}{2pt}}}%
  {\end{list}}
\newcommand\Noted[1]{} 

\definecolor{darkblue}{rgb}{0.0, 0.0, 0.55}
\definecolor{mygreen}{HTML}{ADFF2F}
\definecolor{mylightgray}{gray}{0.8}

  {\begin{itemize}
	[leftmargin=0cm,
		itemindent=.3cm,
		labelwidth=\itemindent,
		labelsep=0pt,
		parsep=0.0pt,
		topsep=0.5pt,
		itemsep=0.5pt,
		align=left]
  }%
  {\end{itemize}}    

  {\begin{enumerate}
	[leftmargin=.cm,itemindent=.5cm,labelwidth=\itemindent,
		labelsep=0pt,
		parsep=1pt,
		topsep=1pt,
		itemsep=3pt,
		align=left]
  }%
  {\end{enumerate}}    


\makeatletter
\def\@copyrightspace{\relax}
\makeatother

%% file: sec-abstract.tex
\begin{abstract}
    Human memory is inherently prone to forgetting. To address this, multimodal embedding models have been introduced, which transform diverse real-world data into a unified embedding space. These embeddings can be retrieved efficiently, aiding mobile users in recalling past information. However, as model complexity grows, so do its resource demands, leading to reduced throughput and heavy computational requirements that limit mobile device implementation. In this paper, we introduce \sys, a novel on-device multimodal embedding system optimized for resource-limited mobile environments. \sys achieves high-throughput, accurate retrieval by generating coarse-grained embeddings and leveraging query-based filtering for refined retrieval. Experimental results demonstrate that \sys delivers high-quality embeddings with superior throughput, all while operating unobtrusively with minimal memory and energy consumption.
\end{abstract}

%% file: todo.tex



%% file: sec-intro.tex
\section{Introduction}
\input{fig-intro.tex}

Mobile devices are ubiquitous nowadays. 
They capture lots of data in users' daily usage, which are invaluable to making devices intelligent assistants~\cite{li2024personal, de2020intelligent, ahmad2023data, zhang2024llamatouch}.
For example, this data can be used for memory recall, helping users retrieve specific information or moments from the past.
For instance, Microsoft launches a project called Recall that makes a note of everything ever displayed on personal computer for AI-empowered retrospective search~\cite{microsoft-recall}.

However, such data has not been fully utilized, attributed not to how to \textit{store} them, but how to accurately \textit{retrieve} them~\cite{273827}.
Specifically, smartphones have abundant storage (up to 1TB for iPhone 15 Pro) to host the information captured at 24x7, or local network-attached storage (NAS) can help accommodate those data as well; yet there has been a lack of method to efficiently locate the data intended at query time~\cite{de2023pre, izacard-grave-2021-leveraging}.
The fundamental challenge is that data generated on devices is multimodal by nature (e.g., text, image, audio, IMU, etc), which are hard to be accurately retrieved in a user-friendly manner, e.g., through natural language~\cite{li2023cross}.

Fortunately, the recent development of multimodal embedding models (MEM) has shed light on multimodal data retrieval.
For example, CLIP unifies text and image modalities into one embedding space~\cite{radford2021learning}.
ImageBind further extends the functionality to 6 modalities through contrastive learning~\cite{girdhar2023imagebind}.
At architecture level, those models primarily consist of multi-layer transformer encoders~\cite{vaswani2017attention}.

In general, MEMs will catelyze two novel, exciting types of mobile applications as shown in Figure~\ref{fig:intro}:
(1) \textit{cross-modality searching}, which allows users to retrieve data in any modality with user-friendly interface, e.g., language;
(2) \textit{retrieval-augmented LLM generation}, which first identifies the relevant multimodal data (e.g., a picture) in a historical database with user prompt, and uses it to enhance the LLM generation quality, e.g., ``in the picture I took for my kid yesterday, is she wearing a blue skirt or yellow?''.



This work addresses the emerging scenario of \textit{on-device multimodal embedding}, where MEMs operate on local devices to embed continuous data streams~\cite{xu2024empowering, li2024large, yuan2024mobile, xu2024survey}. The local generation of embeddings is motivated by user privacy concerns, since MEMs can greatly expand the usage of device data, including screen UIs, recorded voices, etc.
Offloading such information to the cloud may expose it to unauthorized access.
For instance, it was revealed that Apple Siri had been eavesdropping on uploaded user conversations to enhance their public voice assistant model~\cite{siri-listen}. 
With cloud-based MEMs, users risk comprehensive life surveillance, with no way to verify.

\paragraph{Cost of on-device MEMs.}
Despite MEM is generalizable to various downstream tasks~\cite{girdhar2023imagebind, fei2022towards, li2024multimodal, chameleonteam2024chameleonmixedmodalearlyfusionfoundation}, it comes at a cost of resource intensity.
Specifically, our pilot experiments in $\S$\ref{sec:bkgnd-preliminary} identify two key obstacles towards on-device multimodal embedding:
(1) \textit{Low embedding throughput.} 
    It takes dozens of seconds for billion-sized MEMs to embed a single image, which is significantly slower than the rate at which mobile devices generate data.
    As a result, even if the CPU runs continuously, only 20\% of daily information can be embedded.
    (2) \textit{High energy consumption.} The slow inference speed, combined with the immense computing power required, results in extremely high energy consumption.
    Embedding data from applications consumes even more energy than running the applications themselves.
    As a result, the battery life of mobile devices is significantly reduced, often to less than 2 hours.
    Even if the embedding process is batched and executed offline (e.g., when the device is idle), its substantial resource demands still hinder practical deployment.


\paragraph{Our response: \sys.}
We present \sys, the first-of-its-kind efficient on-device multimodal embedding system.
The key idea behind \sys is \textit{coarse-grained embedding}, built upon the early-exiting technique.
Early exiting is a well-known approach in both traditional CNNs~\cite{teerapittayanon2016branchynet,wang2019dynexit,cui2022dvabatch,fei2022deecap} and recent language models~\cite{hamed2024multimodal,gromov2024unreasonable,elhoushi2024layer}, but we found it is rarely integrated with MEMs or adapted for on-device MEMs (\S\ref{sec:design-overview}).
In this work, we revisit the early-exiting technique from the perspective of on-device MEMs and apply it to these models.
We refer to the embeddings generated by the exited MEMs as coarse-grained embeddings, which are used to filter out the most likely candidates during retrieval queries.
These candidates are then further refined at query time to finalize accurate retrieval.



\paragraph{Challenges of early exiting in MEMs.}
While early-exiting-based coarse-grained embedding avoids full model execution during memorization, three key system challenges remain:
\textit{(1) Low parallelism.} Early exiting does not work well with batching, as all samples in a batch must exit before a new batch can be processed~\cite{teerapittayanon2016branchynet}. 
This further exacerbates the throughput issue on mobile devices with limited computational power.
\textit{(2) Limited benefits.} 
Only 30\% of computation can be saved by exiting early in MEMs (\S\ref{sec:design-overview}), even with well-trained exiting heads.
MEMs need to further reduce the number of layers required to predict each token.
\textit{(3) Performance degradation.} Some samples will inevitably exit too early, which is especially problematic in MEMs. 
Incorrect embeddings disrupt the unified embedding space, leading to unbalanced distributions and inaccurate retrieval.

In this work, we aim to address these challenges by proposing \sys, a system designed to efficiently generate precise embeddings for multiple modalities in the background through three hardware-algorithm co-designs.

\textbf{Data-aware pre-exit predictor} (\S\ref{sec:design-predict}):
To enhance data parallelism, the key is to estimate exit points early for better system scheduling.
From a data perspective, information content varies across messages.
Thus, we introduce a data-aware pre-exit predictor for coarse-grained embeddings. 
Unlike traditional methods that determine exit values after each branch computation, our approach uses a unified, lightweight early-exit predictor model applicable to all modalities.
This capability facilitates efficient batching and pipeline execution, significantly improving encoding throughput.

\textbf{Progressive LoRA healing} (\S\ref{sec:design-healing}):
To ensure high retrieval performance even with earlier exits, we retrofit low-rank adaptation (LoRA)~\cite{hu2021lora}, a popular parameter-efficient fine-tuning method, to optimize the exited coarse-grained embeddings.
Traditional LoRA fine-tuning is not designed for early exiting because it requires separate tuning for each exit branch. 
This results in duplicated forward passes for multiple exit branches during a single inference procedure.
Instead of retraining the entire LoRA from scratch, we propose sharing previously tuned LoRA weights for each newly added layer. 
Intermediate results can be cached and reused in subsequent forward passes, improving embedding throughput.
We also observe that the optimal tuning step, which determines how many LoRA layers are tuned together, varies significantly.
To address this, we design a dynamic step scheduler, orchestrated with the data-aware pre-exit predictor, prioritizing healing at exit points with the most sample exits.


\textbf{Speculative fine-grained retrieval} (\S\ref{sec:design-query}):
To enhance search performance, coarse-grained embeddings with high potential require refinement. 
\sys implements a speculative retrieval mechanism that refines these embeddings during query time. 
The remaining layers of the exited MEMs serve as a live encoder to refine coarse-grained embeddings and finalize retrieval. 
To ensure balanced retrieval, the first filtering round uses query embeddings of various granularities, which are generated by different exits in the model. 
The top candidates from each granularity are then selected for a second round of matching, ensuring accurate final retrieval. 
This approach offloads fine-grained embedding refinement to query time, where the search objective is clear, enabling rapid memorization and precise recall.

\paragraph{Results}
We implement \sys atop ImageBind, a widely used MEM pretrained by Meta~\cite{girdhar2023imagebind}.
We evaluate \sys on three devices: NVIDIA ORIN~\cite{orin}, Raspberry Pi 4B~\cite{rpi4b}, and a flagship smartphone with Qualcomm Snapdragon 8Gen3~\cite{rn12t}.
\sys delivers an average 14.9$\times$ improvement in throughput and 13.1$\times$ reduction in energy consumption compared to the original MEM.
It efficiently embeds the vast majority of daily usage data, minimizing battery drain.
Our key designs are essential for achieving these improvements while maintaining high accuracy.
The combined system incurs less than 5\% relative accuracy loss compared to the full-sized MEM, with query delays under 5 seconds.
Additionally, we conduct a case study using recent Twitter data and a user study on mobile application traces, demonstrating the practicality of \sys in real-world scenarios.

\paragraph{Contributions.}
We make the following contributions:
\begin{itemize}
    \item We prototype the first MEM-empowered mobile search service architecture. Through user studies and pilot experiments, we identify challenges related to low embedding throughput and high energy consumption\footnote{Codebases and collected trace data will be made public after acceptance.}.
    \item We introduce \sys, an efficient on-device multimodal embedding system that addresses these challenges. \sys incorporates three novel techniques: preemptive exit for dynamic execution scheduling, progressive model healing for cache optimization, and speculative retrieval to correct premature exits.
    \item Extensive experiments demonstrate that \sys significantly improves throughput and reduces energy consumption while maintaining search performance, making it practical for modern mobile devices.
\end{itemize}


%% file: fig-intro.tex
\begin{figure}[t]
	\centering
	 \includegraphics[width=0.5\textwidth]{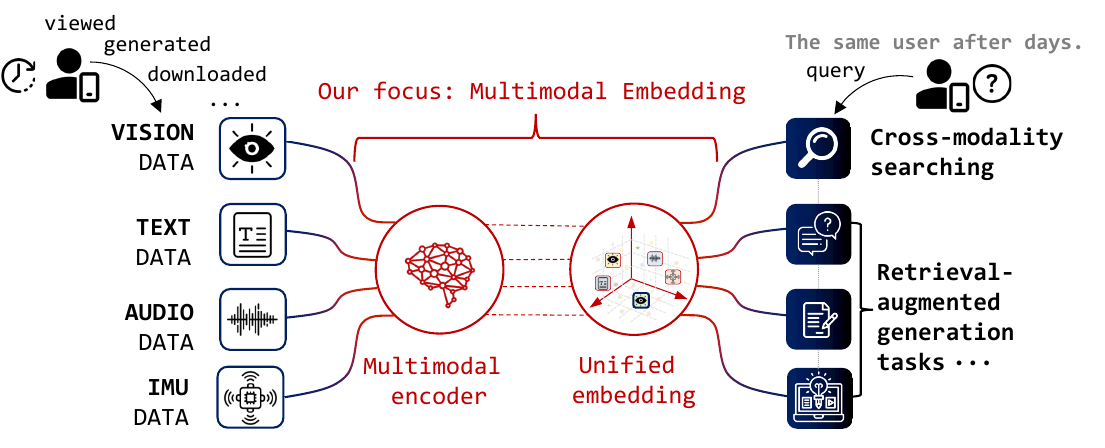}
	 \vspace{-20pt}
	\caption{MEM workflow and its application.
	} 
	\vspace{-20pt}
	\label{fig:intro}
\end{figure}

%% file: sec-bkgnd.tex
\section{Background and Motivations}
\label{sec:bkgnd}


\subsection{Multimodal Embedding}


\paragraph{Unified multimodal embedding}
Embedding was initially proposed to vectorize text data for understanding similarities between different texts~\cite{lin2017structured}.
Large language models use embedding layers to generate text embeddings~\cite{devlin2018bert,vaswani2017attention}.
Similarly, vision, audio, and sensor data can also be transformed into vectorized embeddings~\cite{dosovitskiy2020image,radford2023robust,hong2024spectralgpt}.
However, embedding methods focused on a single modality cannot access information across different modalities due to the gap between their embedding spaces.

To bridge this gap, multimodal embedding models (MEMs) have been developed to unify different modalities into a single embedding space, enhancing the model's ability to understand and bind multimodal inputs.
CLIP~\cite{radford2021learning} aligns text and vision by jointly training on image-text pairs, using contrastive learning to map both modalities into a shared space while maintaining their distinction through a dual-tower architecture.
ImageBind~\cite{girdhar2023imagebind} extends this to align six modalities, including text, vision, audio, depth, thermal, and IMU readings.
Each modality is processed by a separate encoder, and the embeddings are fused in a multimodal head to generate a unified embedding.
ImageBind demonstrates strong zero-shot classification and retrieval performance across these modalities, matching or outperforming single-modality models.
This is achieved through training on large-scale multimodal data.

\paragraph{Multimodal mobile applications}
MEMs optimize alignment between high-quality representations across modalities.
As such, multimodal information can be composed to enable a rich variety of mobile context-aware applications.
For example, MEMs could embed visual, audio, text and sensor data experienced on a mobile device into a personalized memory palace~\cite{fassbender2006virtual,huttner2019immersive}. 
Whenever users want to recall a specific moment or items, they can query the memory palace with a multimodal query, and the system will retrieve the most relevant items.
MEMs can also facilitate mobile agents to iteract with users in a more human-like manner~\cite{li2024personal,yang2023appagent,liu2020indoor}.





\paragraph{On-device Multimodal Embedding}
Data for embedding is continuously sourced from end users and is often private and sensitive.
Evidence suggests that cloud service providers may be curious about uploaded data to improve their services~\cite{siri-listen}, and database leaks and breaches pose significant threats~\cite{apple-leakage}.
Conducting embedding locally prevents the need to upload daily viewed, sensed, or heard data to the cloud, offering strong privacy protection.
From the cloud perspective, a single user views over 6,000 images per day, according to our user study (\S\ref{sec:bkgnd-preliminary}), requiring approximately 1065.6KJ of energy and 0.8 GPU hours.
For 1 billion daily active users, cloud providers would need 1.1 TWh of energy and 0.8M GPU hours daily, costing over \$100 million per day.
On-device multimodal embedding shifts this cost to end users, making the service more practical to deploy.

\subsection{MEM-empowered Search Service}
\label{sec:bkgnd-sys-model}
\input{fig-design-application.tex}

As shown in Figure~\ref{fig:design-application}, we prototype an on-device MEM-powered search service to embed multimodal streaming data for future retrieval, functioning like a memory palace~\cite{fassbender2006virtual}.
We specifically target mobile devices, including smartphones and IoT devices with similar computing capabilities.
These devices have usable but weaker processing units compared to cloud servers, with limited battery and memory available for long-term background processes~\cite{os-killer}.


From the device perspective, the service consists of two runtimes:

\begin{itemize}
    \item \textbf{Embedding runtime (Offline remembering in the backend)} continuously detects and stores newly generated multimodal content, such as downloaded images, scanned texts, listened-to audio, and logged IMU sensor data.
    Each item is processed layer by layer through MEMs\footnote{Deep learning models are often too large for mobile devices, leading to inference processes being terminated by the OS. Current mobile inference engines provide layerwise execution to support large models~\cite{mllm24, Ni_ncnn_2017}.}, generating 1024-dimensional embeddings in a unified space.
    
    \item \textbf{Query runtime (Online recall in the frontend)} is triggered when the user searches for a specific item or performs other tasks based on search results. 
    To retrieve relevant items, the query embedding is compared with stored embeddings to find the most similar matches.
    If the raw data corresponding to the matched embeddings aligns with the query intent, the query is tagged as successful.
\end{itemize}


System developers prepare the embedding model offline, typically by fine-tuning with powerful cloud GPUs, using widely-used pretrained multimodal embedding models~\cite{girdhar2023imagebind,radford2021learning}.
They define the expected offline costs and online performance for each application by configuring system hyperparameters before deployment, as shown in Figure~\ref{fig:design-application}.

\subsection{Preliminary Measurements}
\label{sec:bkgnd-preliminary}

\input{fig-motivation-trace.tex}
We conducted a user study to collect viewed images from daily mobile applications used by 8 volunteers, aged 20 to 52, over the course of a week.
To achieve this, we developed an Android application with accessibility services~\cite{accessibility} to detect and store newly appeared visual content\footnote{Images are hashed to include only new content. Images smaller than 100KB are excluded to avoid capturing icons and minor system elements.}.
One collected trace is shown as an example in Figure~\ref{fig:motivation-trace}.


\input{fig-motivation-sample.tex}
\paragraph{Observation: MEMs are contextually expressive.}

All images and corresponding texts are collected and embedded using ImageBind~\cite{girdhar2023imagebind}.
By aligning multimodal embeddings into a common space, ImageBind can effectively retrieve semantically relevant content from different modalities using human-friendly input formats.
For example, as shown in Figure~\ref{fig:motivation-sample}, the sound of fireworks retrieves images of fireworks from the albums and their corresponding textual notes with high confidence.
A rigorous numerical analysis across various tasks will be presented in \S\ref{sec:eval}.



\paragraph{Challenge: MEMs are resource-intensive.}
\input{fig-motivation-throughput.tex}

To assess the cost of on-device embedding, we ran ImageBind inference on four different mobile devices, ranging from development boards to commodity smartphones.

\textbf{Huge workloads and low throughput.}
Despite their contextually expressive capabilities, the embedding speed is too slow to keep pace with the figures generated by applications.
As shown in Figure~\ref{fig:motivations-cost}, on all CPU-based devices, the encoding speed is insufficient for real-time application use.
Over a full day of usage, the speed is only sufficient to embed 20\% of the figures generated by applications, requiring more than 100 hours to process all figures from a single day.
Even with a GPU, Jetson NANO~\cite{kurniawan2021introduction} struggles to handle an entertainment task generating 36.3 images per minute.
The only exception is the NVIDIA ORIN~\cite{orin}, which performs comparably to a cloud server using an NVIDIA A40~\cite{orin-capacity}.
However, continuously running the CPU or GPU on mobile devices is impractical due to battery depletion.

\textbf{Battery depletion.}
The heavy embedding workloads and low throughput significantly strain battery life.
Continuous embedding drains the battery even faster than running the app itself.
To illustrate, we used ImageBind to continuously embed figures from daily apps.
As shown in Figure~\ref{fig:motivations-battery}, the embedding process consumes more energy than the apps themselves. 
For example, even when quantized to INT4, MEMs consume 1.8$\times$ more energy than gaming.
We also measured GPU energy consumption on an NVIDIA ORIN\footnote{Current mobile inference engines cannot effectively utilize GPUs for MEM execution~\cite{cai2021towards, li2024large,mllm24}.}.
While GPUs process data faster, they consume more energy than CPUs, making them unsuitable for long-term embedding in the current MEM design.

%% file: fig-design-application.tex
\begin{figure}[t]
	\centering
	 \includegraphics[width=0.38\textwidth]{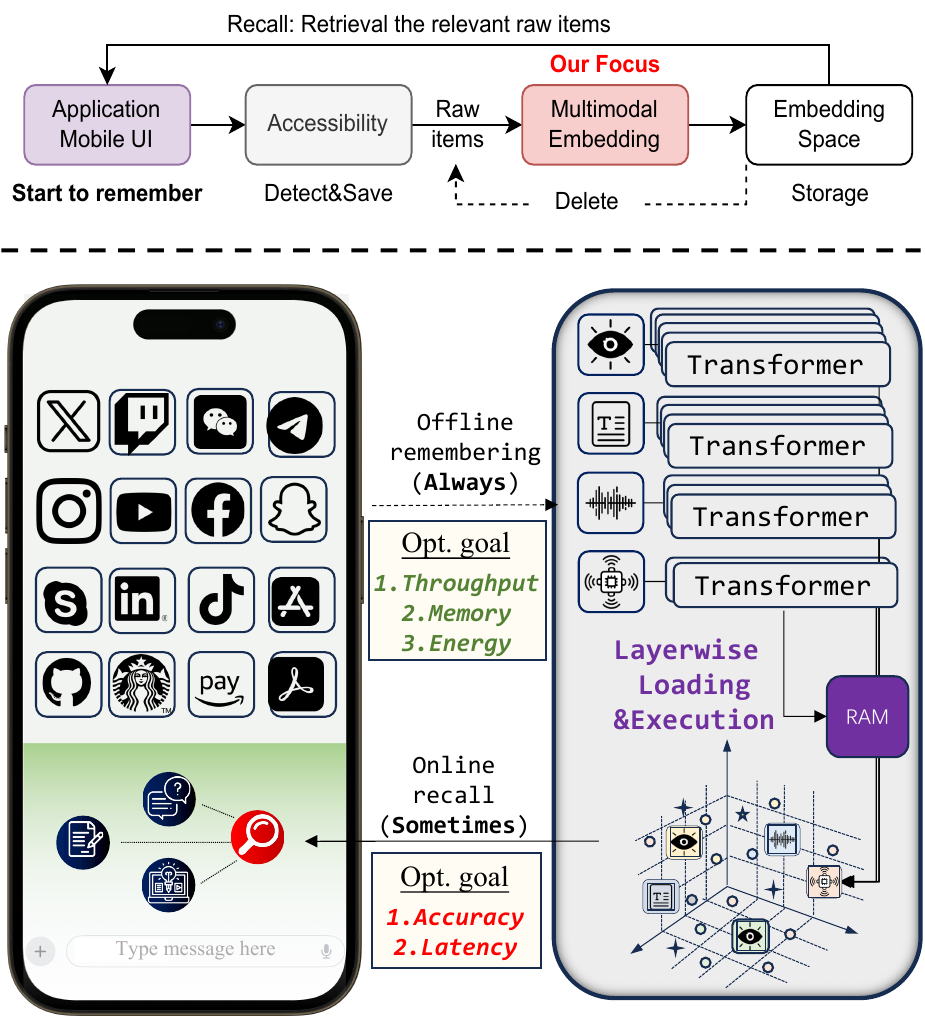}
	\caption{\service provides system-level service that remembers daily mobile interaction for  recalling.
	} 
	\vspace{-15pt}
	\label{fig:design-application}
\end{figure}

%% file: fig-motivation-trace.tex
\begin{figure}[t]
	\centering
	 \includegraphics[width=0.48\textwidth]{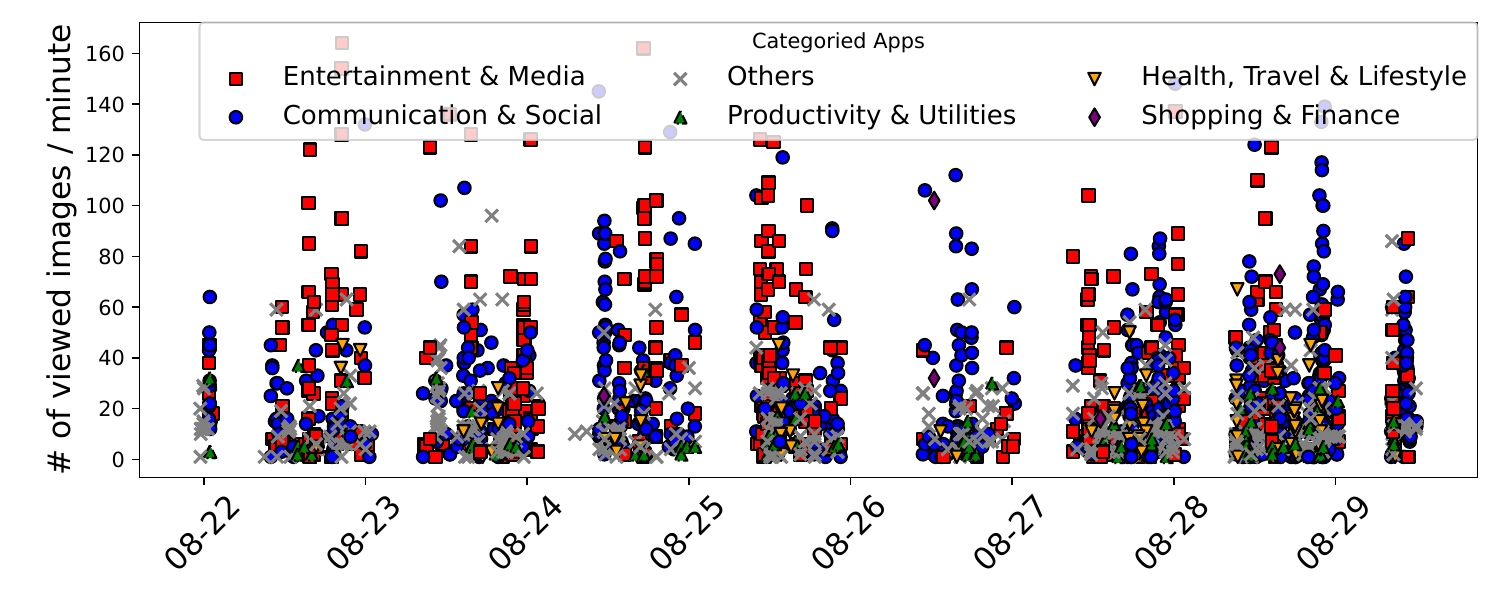}
	 \vspace{-20pt}
	\caption{Viewed image trace of mobile users.} 
	\vspace{-10pt}
	\label{fig:motivation-trace}
\end{figure}

%% file: fig-motivation-sample.tex
\begin{figure}[t]
	\centering
	 \includegraphics[width=0.48\textwidth]{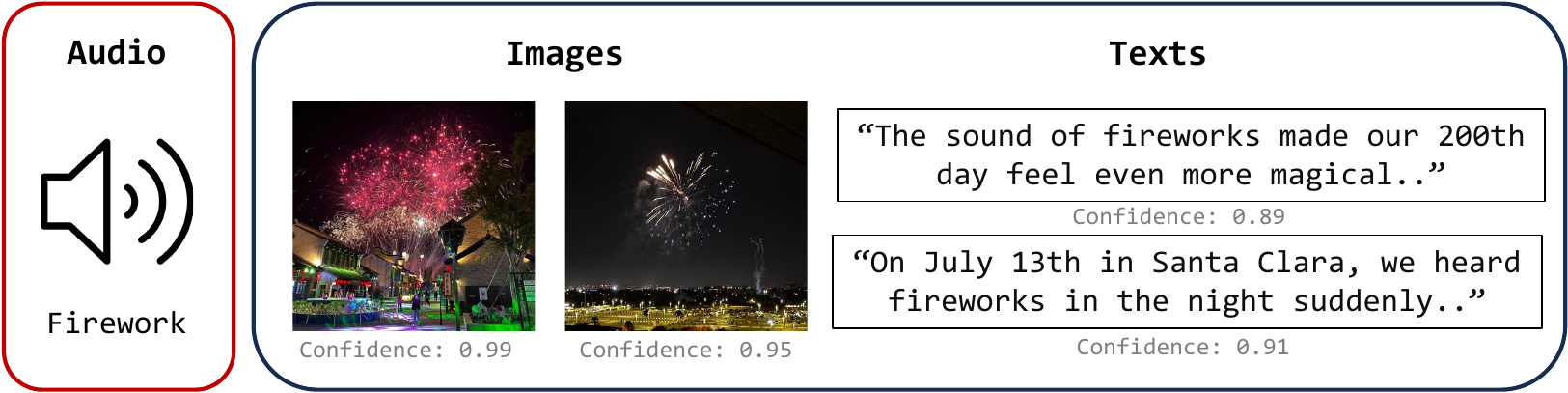}
	 \vspace{-20pt}
	\caption{Demo of cross-modal retrieval.} 
	\vspace{-10pt}
	\label{fig:motivation-sample}
\end{figure}

%% file: fig-motivation-throughput.tex
\begin{figure}[t]
    \centering
    \begin{minipage}[b]{0.24\textwidth}
        \centering
        \includegraphics[width=0.95\textwidth]{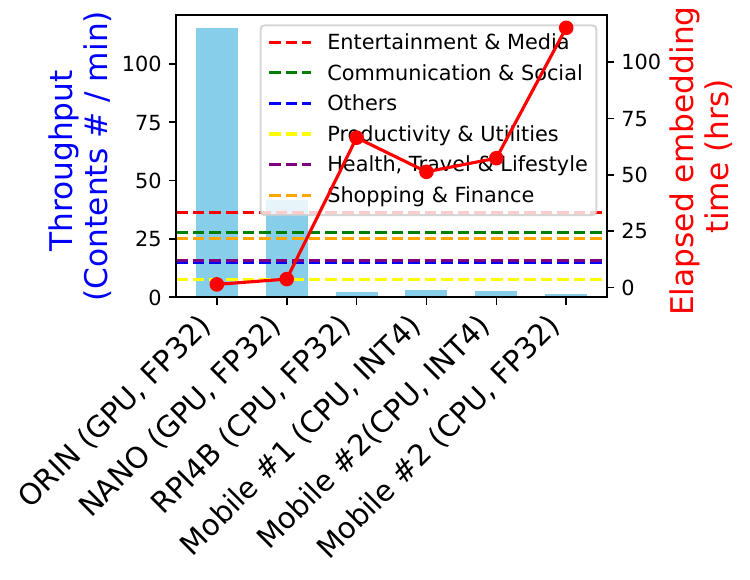}
        \subcaption{Throughput} 
		\label{fig:motivations-cost}
    \end{minipage}
    ~
    \begin{minipage}[b]{0.24\textwidth}
        \centering
        \includegraphics[width=0.98\textwidth]{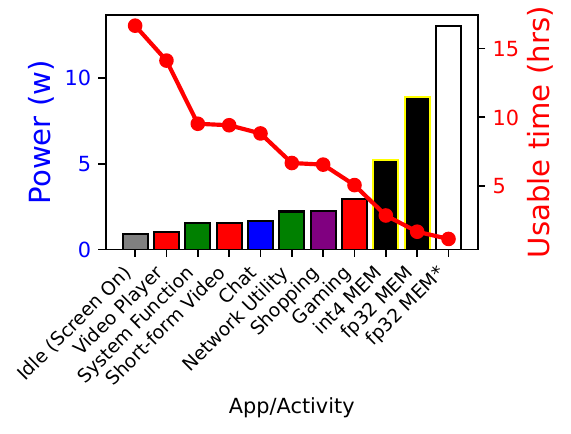}
	\subcaption{Battery} 

	\label{fig:motivations-battery}
    \end{minipage}
    	\vspace{-15pt}
    \caption{Throughput and energy issues. (a) MEM inference speed across different devices compared to the average image generation speed of various mobile applications. (b) MEMs rapidly drain the battery of the test mobile phone. * indicates GPU inference power consumption of the Jetson ORIN.
    }
    \vspace{-10pt}
    \label{fig:motivations-energy}
\end{figure}

%% file: sec-design.tex
\section{Design}
\input{fig-design-overview.tex}



\subsection{\sys Overview}
\label{sec:design-overview}

In this work, we develop \sys, an efficient on-device multimodal embedding system to address the challenges outlined above.
\sys is designed to minimize embedding energy costs and query latency while maximizing throughput and achieving near state-of-the-art retrieval accuracy.
Additionally, \sys shall integrate easily into off-the-shelf mobile applications to enhance user experience without requiring complex hardware modifications.
Lastly, \sys aims to be both versatile and transferable across a wide range of tasks.
To achieve these goals, we leverage early exit, a widely studied optimization technique, as the backbone of our system.

\paragraph{Key building block: early exit} terminates the computation of a deep neural network at an intermediate layer based on prediction confidence.
Typically, a prediction head is introduced at the end of each layer to serve as a separate exit branch, allowing samples to be correctly classified at the earliest possible layer.

We choose early exit as the backbone of \sys because it aligns with our design principles:
(1) Early exit is mobile hardware-friendly: it requires no sparsification kernel compilation and integrates easily into existing multimodal embedding applications. Most mobile devices do not fully support advanced sparsification or quantization optimizations, providing little to no benefit during inference~\cite{lu2021sanger, kim2023full, sun2022speformer, armeniakos2022hardware, wang2021spatten}.
(2) Early exit preserves the raw structure of MEMs, maintaining their generalization capacity while bypassing only downstream alignment.
Additionally, early exit is caching-friendly, as the top layers share the same bottom weights with the exited layers, allowing intermediate activations to be reused and reducing duplicated computations. Other techniques like pruning and quantization cannot fully leverage the intermediate computation of coarse-grained embeddings.
This reduction is crucial for \sys, as it eliminates redundant forward passes, accelerating both embedding and query phases, which we discuss in detail in \S\ref{sec:design-query}.


\paragraph{Simplified workflow:}
As shown in Figure~\ref{fig:design-overview}, \sys provides a memory encoder for clients to build coarse-grained embeddings offline, while the rest of the model functions as a live encoder for precise online retrieval.
(1) \textit{System developer preparation:} Developers first refine widely-used pretrained multimodal models to reduce the number of layers needed for token prediction (\S\ref{sec:design-healing}).
The refined model is then deployed to mobile devices for offline embedding.
(2) \textit{Client offline embedding:} Users employ part of the memory encoder to build superficial embeddings for pre-exit prediction  (\S\ref{sec:design-predict}). After pre-exit, samples with the same exits are batched and processed layer by layer through pipeline scheduling to generate coarse-grained embeddings.
(3) \textit{Client online query:} During the query phase, the query is embedded for matching. Likely candidates are filtered and refined from the coarse-grained embeddings, which are then matched with the query embedding to finalize retrieval (\S\ref{sec:design-query}).

In short, we offload the full-sized embedding cost to the query phase, which is infrequent and carries precise retrieval information~\cite{de2023pre}. 
This mirrors the human brain, which retains key information in long-term memory and recalls details only when necessary~\cite{banikowski1999strategies}. 
Retrieval accuracy and latency are sacrificed within acceptable limits to significantly reduce embedding costs, as demonstrated in \S\ref{sec:eval}.

\paragraph{Unique challenges introduced by early exit:}
While early exit reduces computational load, its application in mobile MEMs introduces several unique challenges:
(1) \textit{Low parallelism:} Early exit is incompatible with batching, as all samples in a batch must exit before processing the next~\cite{teerapittayanon2016branchynet}. 
    This significantly reduces throughput on mobile devices with limited computational resources. Without batching, it is also harder to amortize loading costs, further slowing layer-wise inference.
(2) \textit{Limited benefits:} MEMs are not naturally designed for early prediction and tend to distribute computation across all layers. For instance, ImageBind's 32-layer vision module requires an average of 21.4 layers to process data, limiting computation savings to 33.1\%.
    MEMs need to reduce the layers required for token prediction and minimize computational resources spent on hesitant or fluctuating predictions.
(3) \textit{Performance degradation:} Despite thorough training of exit branches and predictors, some samples may exit too early, leading to degraded search performance. 
    This is especially problematic in MEMs, where incorrect embeddings can disrupt the unified embedding space, causing unbalanced distributions and inaccurate retrieval.

\input{sec-design-predict.tex}

\input{sec-design-lora.tex}

\input{sec-design-query.tex}

%% file: fig-design-overview.tex
\begin{figure*}[htbp]
	\centering
	 \includegraphics[width=0.95\textwidth]{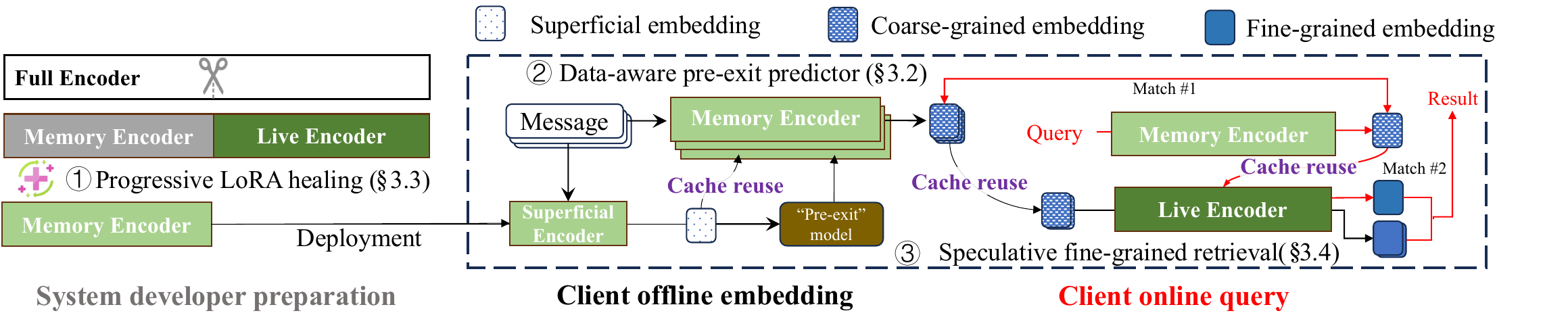}
	 \vspace{-10pt}
	\caption{Overview of \sys.} 
	\label{fig:design-overview}
\end{figure*}

%% file: sec-design-predict.tex
\subsection{Data-aware Pre-exit Predictor}
\label{sec:design-predict}

\input{fig-design-predictor.tex}

Traditionally, most early-exit methods decide whether to exit at the end of each branch computation~\cite{laskaridis2021adaptive,teerapittayanon2016branchynet,elhoushi2024layer}.
This approach limits hardware acceleration and batching, as exit points vary by data, leading to inconsistent workloads within batches and memory fragmentation~\cite{teerapittayanon2016branchynet,li2023predictive,wang2019dynexit}.
Although some predictive models for CNNs~\cite{wang2019dynexit} predict exit values in advance, they cannot scale to MEMs due to their convolution-specific design.
In this work, we propose a unified, lightweight early-exit predictor model for all modalities, derived from intermediate data embeddings.
The data-aware pre-exit predictor preemptively decides the exit point for MEMs, enabling batch scheduling for better parallelism and helping to amortize and hide loading time.

\input{fig-design-predict-data.tex}

\paragraph{Data-aware coarse-grained embedding granularity}
Different data contains varying amounts of information content.
Unlike previous work that defines predictive models manually, we propose using intermediate embeddings to predict the exit value without supervision.
First, we build the fine-grained embedding $\mathtt{F}_{x}$ for each data point $x \in \mathtt{X}$ as a proxy query label.
Next, we feed the input into the pre-trained MEM layer by layer, obtaining a set of coarse-grained embeddings $\mathtt{C}_{x}^{i}$ at different granularities $i \in \text{range}(\text{layers})$.
We then measure the similarity between the fine-grained and coarse-grained embeddings.
When the similarity between $\mathtt{F}_{x}$ and $\mathtt{C}_{x}^{i}$ becomes the largest among $\mathtt{F}_{x}$ and $\mathtt{C}_{\mathtt{X}}^{i}$. query retrieves $\mathtt{C}_{x}^{i}$ from $\mathtt{C}_{X}^{i}$ successfully.
We mark it as a valid embedding exit.
The intermediate embeddings are fed into the predictor model, and an MLP model is trained to predict its exit value.
This method outperforms fixed early-exit baselines, as will be shown in \S\ref{sec:eval-ablation}.



\paragraph{Batch-friendly and pipeline execution}
As shown in Figure~\ref{fig:design-pre-exit-workflow}, with the data-aware pre-exit predictor, we can predict the exit value before embedding, enabling efficient batching of input data.
In addition to early-exit-specific batching, we propose pipelining the layer-by-layer encoding process, where loading and embedding are conducted simultaneously.

\input{alg-predict.tex}

\paragraph{Pre-exit Predictor in detail}
We summarize the use of the pre-exit predictor in Algorithm~\ref{alg:predict}.
First, we load Layer$_i$ and encode all input data as a batch, while Layer$_{i+1}$ is loaded concurrently to minimize loading time.
This process iterates until all $N$ layers are loaded.
Next, we feed the intermediate embeddings (i.e., superficial embeddings) to the predictor model.
Data are then batched according to the predicted exit values.
These steps are repeated for each batch until all data reach their predicted exits.

\paragraph{Pre-exit predictor cost}
Training the predictor is efficient, requiring only tens of iterations on hundreds of samples, taking just a few minutes on a single GPU.
The trained predictor is lightweight, with a memory footprint of around 1MB.
The main concern is the cost of computing the superficial embedding.
Fortunately, this embedding can be reused for subsequent coarse-grained embeddings, as discussed in \S\ref{sec:design-query}.


\paragraph{Micro Experiments}
are conducted to demonstrate the effectiveness of the pre-exit predictor.
As shown in Figure~\ref{fig:design-dynamic-perf}, prediction accuracy improves with the increase of superficial embedding layers.
As indicated by Figure~\ref{fig:design-dynamic-rationale}, most samples require the complexity of more than 7 layers.
With \(N = 7\), the predicted accuracy is 85\%, the average predicted layer is 15.5, and the average actual layer is 16.5.
An interesting finding is that as the intermediate embeddings are fed layer by layer, the deeper the layers, the more accurately the predictor model can determine the exit value.
This improvement occurs because deeper layer embeddings are more discriminative and better suited for predicting the final embedding.

%% file: fig-design-predictor.tex
\begin{figure}[t]
	\centering
	 \includegraphics[width=0.48\textwidth]{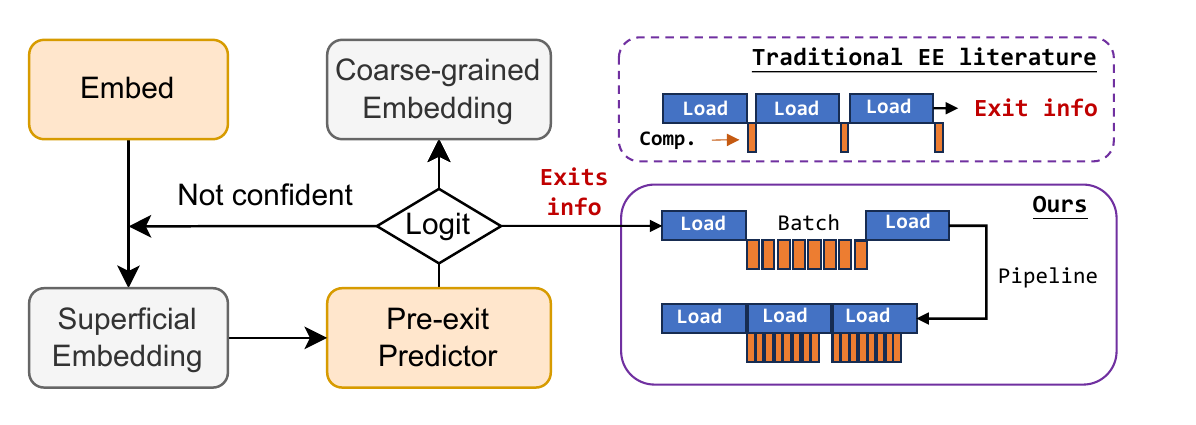}
	 \vspace{-15pt}
	\caption{Data-aware pre-exit workflow.} 
	\label{fig:design-pre-exit-workflow}
\end{figure}

%% file: fig-design-predict-data.tex
\begin{figure}[t]
    \centering
    \begin{minipage}[b]{0.24\textwidth}
        \centering
        \includegraphics[width=0.8\textwidth]{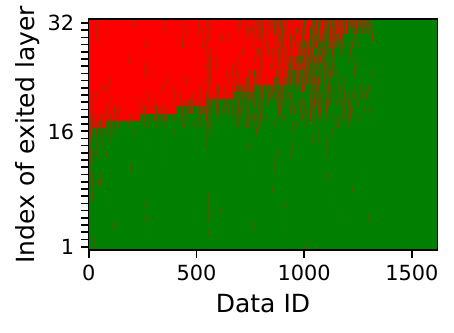}
        \vspace{-3pt}   
        \subcaption{Exit stastics.} 
		\label{fig:design-dynamic-rationale}
    \end{minipage}
    ~
    \begin{minipage}[b]{0.24\textwidth}
        \centering
        \includegraphics[width=0.9\textwidth]{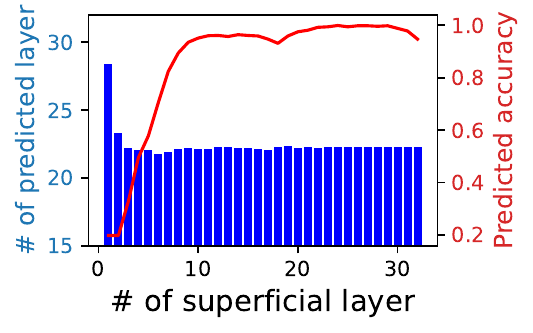}
        \subcaption{Predictor performance.
        } 
		\label{fig:design-dynamic-perf}
    \end{minipage}
    \caption{(a) Each dataset has a different optimal exit point. 
    (b) The data-aware predictor is well-trained to assign the appropriate exit for each sample.
    }
    \vspace{-10pt}
    \label{fig:design-predict}
\end{figure}

%% file: alg-predict.tex
\begin{algorithm}[t]
	\footnotesize
	\SetAlgoNoEnd
  \SetKwProg{Fn}{Function}{~}{end}
  \SetKwData{Left}{left}\SetKwData{This}{this}\SetKwData{Up}{up}
  \SetKwFunction{MIN}{MIN}\SetKwFunction{MAX}{MAX}\SetKwFunction{LENGTH}{LENGTH}
  \SetKwInOut{Input}{input}\SetKwInOut{Output}{output}\SetKwInOut{Variable}{Variable}

  \Input{
      Superficial Embedding Layer $N$;\\
      Predict model $\phi_{S}$;\\
      Burst-in Streaming Input, $\mathbb{X} $.
    }

  \Output{
      Embedding, $\mathbb{E}$.
    }
  \BlankLine

  \Fn{Data-aware\_Coarse-grained\_Embedding($N$, $\phi_{S}$, $\mathbb{X}$):}{

    Embedding $\leftarrow$ Batched\_Layerwise\_Encoding(0, $N$, $\mathbb{X}$);\\
    Predicted Exit $e$ $\leftarrow$ $\phi_{S}(\mathbb{E})$;\\
    Group $\mathbb{X}$ into $\mathbb{X}^{e}$ with the same exit seperately;\\
    \ForAll{$\mathbb{X}^{e}$}{
      Embedding $\leftarrow$ Batched\_Layerwise\_Encoding($N$, $e$, $\mathbb{X}^{e}$);\\
      }
    Store Embedding $\mathbb{E}$ in the disk.
    }

  \Fn{Batched\_Layerwise\_Encoding($i$, $j$, $\mathbb{X}$):}{
    $\mathbb{X}_{B}$ $\leftarrow$ Batching $\mathbb{X}$;\\
    \ForAll{$\mathbb{X}_{B}$}{
      \While{$i$<$j$}
      { 
        Encode $\mathbb{X}^{i}_{b}$; load layer $i$+1 concurrently;\\
      }
    }
    Embedding $\leftarrow$ $Postprocessor$(Intermediate results);\\
    return Embedding.
  }

  \caption{Our Pre-exit Predictor}\label{alg:predict}
\end{algorithm}

%% file: sec-design-lora.tex
\subsection{Progressive LoRA Healing}
\label{sec:design-healing}

Original MEMs are not designed for early exit, as they tend to distribute computation across all layers.
As a result, most data requires many layers before exiting.
We propose a progressive LoRA approach to heal the model, reducing the number of layers needed for each token.


\input{fig-design-plora-reuse.tex}
\paragraph{Parameter-efficient LoRA Healing}
Previous early-exit healing approaches~\cite{gromov2024unreasonable} use the parameter-efficient fine-tuning method, LoRA~\cite{hu2021lora}, to distill knowledge into lower layers, reducing the number of layers required for each token.
Naive LoRA tuning fine-tunes a separate LoRA suite for each early-exit layer.
For instance, with 32 exits, 32 LoRA suites are required.
While this ensures good performance, it has a significant drawback: the embedding from layer $n$ cannot be reused to compute the embedding for layer $n+1$.
As illustrated in Figure~\ref{fig:design-plora-reuse}, this occurs because LoRA $l_{n}^{1,\ldots,n}$ for layer $n$ is not the same as the first $n$ layers of LoRA $l_{n+1}^{1,\ldots,n+1}$.
Unlike standard embeddings, which complete all layers sequentially, early-exit methods must check whether each layer is the final one.
If layer $n$'s embedding is incompatible with layer $n+1$, the early-exit method must recompute the embedding for layer $n+1$ from scratch, negating many of the benefits of early exit.

On cloud servers, computation is not a major issue due to their high processing power, and reducing model weights to alleviate I/O pressure is the primary concern.
However, for mobile devices with limited computational power, I/O pressure is less of a concern since they typically serve only one user at a time.

\input{fig-design-plora.tex}
\paragraph{Progressive LoRA healing (P-LoRA)}
\sys proposes a progressive LoRA healing method to address this issue, aiming to use a single LoRA suite for all exits.
To achieve this, we tune the LoRA layer by layer.
For each exit, we tune only the LoRA for the current exit while keeping the previous exits' LoRA fixed.
Since the tunable parameters are fewer than the fixed ones, the healing capacity is weaker compared to using separate LoRA suites, which negatively impacts convergence (i.e., fine-grained embedding) performance, as shown in Figure~\ref{fig:design-plora}.
To mitigate this, instead of tuning one LoRA layer at a time, we progressively tune more LoRA layers at later exits.
Similar to the window size in convolutional layers, we define the number of tuned LoRA layers as the LoRA step.

\paragraph{P-LoRA step decision}
As shown in Figure \ref{fig:design-plora}, the optimal healing step varies across exit layers.
In general, the larger the $n$, the greater the per-step healing capacity, due to the increased number of tunable parameters.
However, if step 4 is applied to all exits, exits 2 and 3 will miss opportunities for healing.
This is acceptable for the top layers, as they already have a strong feature representation from earlier healing.
Larger steps benefit later layers by improving convergence performance.
For smaller exits, earlier features are still weak and require healing at each exit.

To determine the optimal step during training, we use information from the predicted exit statistics.
We set the training step at the pivot of the predicted exit statistics, ensuring that most exits are healed with an appropriate step size.
This approach prioritizes smaller exits, aligning with the heuristic that most data exits occur at earlier layers, which require more focused healing.
At later stages, larger steps enhance fine-grained performance during queries without significantly affecting exit flexibility.


\paragraph{Training Details}
The healing P-LoRA is designed to be parameter-efficient and highly transferable.
Application developers can customize the personalized healing adapter during the testing phase.
During deployment, healing occurs iteratively, and embedding granularity can be updated in real time to better fit the data and synchronize with their representations.
In this work, for simplicity, the embedding granularity predictor was trained on zero-shot embeddings.
The training objective is the fine-grained embedding, not the query embedding.
We leave the output layer untuned to mitigate the dynamic embedding mismatch issue (discussed in \S\ref{sec:design-query}).
Even without the healing P-LoRA, we demonstrate that \sys can still achieve usable retrieval performance, as shown in \S\ref{sec:eval-ablation}.


%% file: fig-design-plora-reuse.tex
\begin{figure}[t]
	\centering
	 \includegraphics[width=0.4\textwidth]{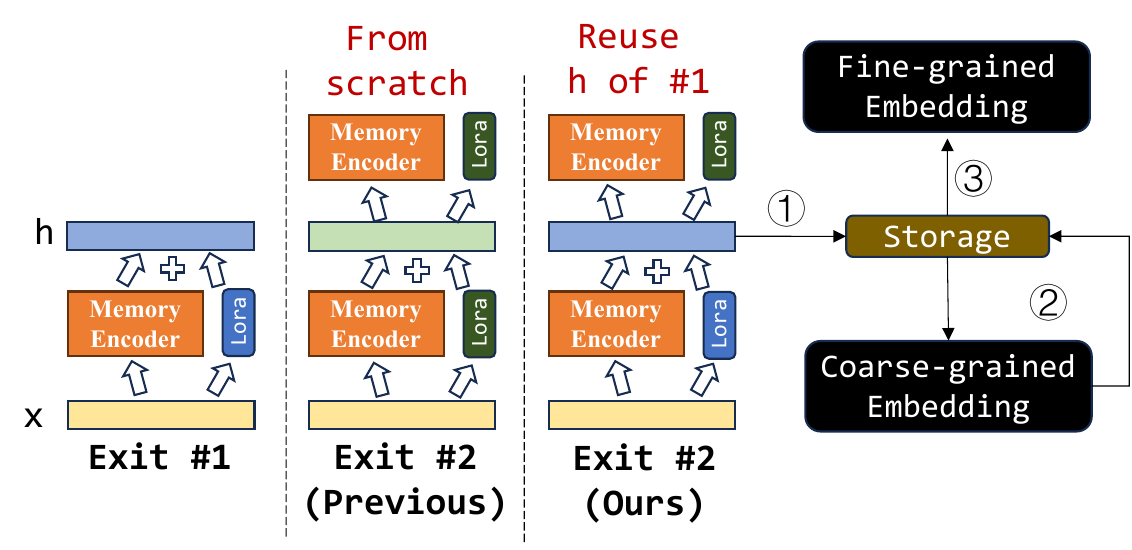}
	 \vspace{-10pt}
\caption{Comparison between Progressive LoRA and previous methods.}
	\vspace{-15pt}
	\label{fig:design-plora-reuse}
\end{figure}

%% file: fig-design-plora.tex
\begin{figure}[t]
	\centering
	 \includegraphics[width=0.48\textwidth]{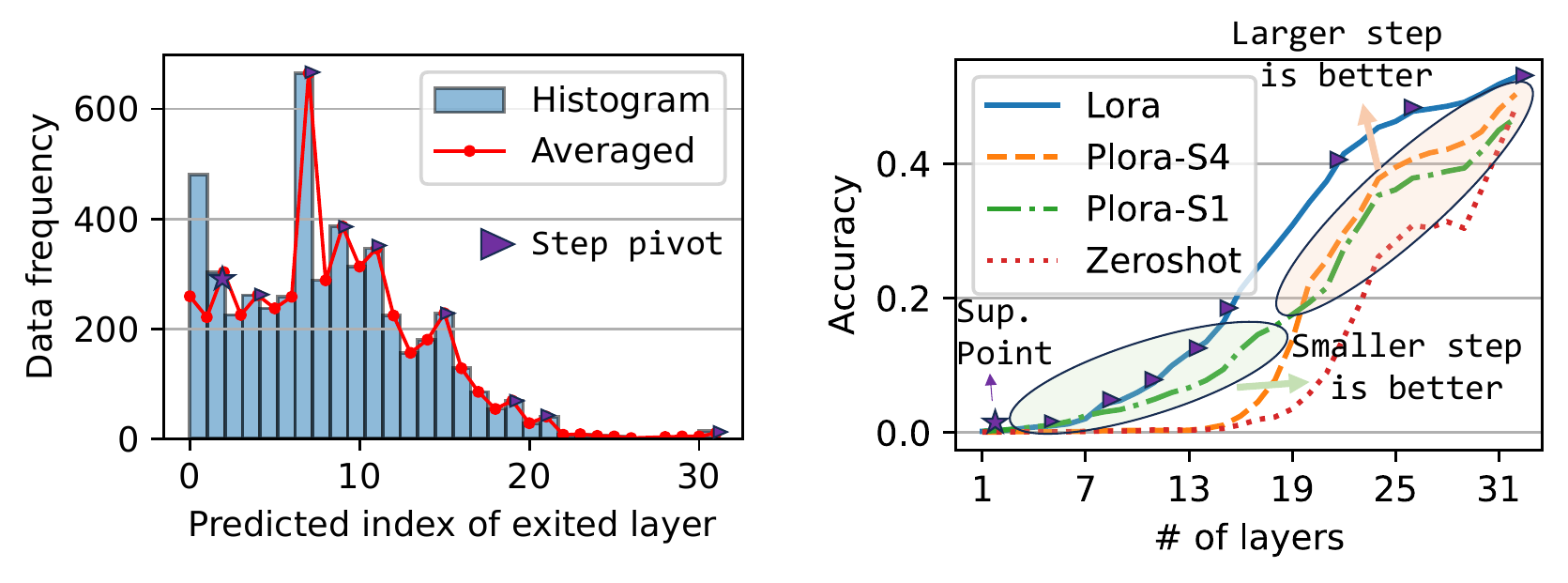}
	 \vspace{-20pt}
\caption{The progressive steps affect tuning performance.}
	\vspace{-15pt}
	\label{fig:design-plora}
\end{figure}

%% file: sec-design-query.tex
\subsection{Speculative Fine-grained Retrieval}
\label{sec:design-query}
\input{fig-design-query.tex}

With coarse-grained embeddings, we can filter out potential candidates.
Further fine-grained embeddings are then processed on these filtered candidates to complete the final retrieval.
However, using the default query embedding with a full-capacity encoder does not achieve precise top-1 retrieval (R@1), as shown in Figure~\ref{fig:design-query}.
This poor performance stems from two unique challenges.

\textit{\# Challenge 1: Reduced embedding capacity.}
Even if we modify the model to predict early and align it with the full embedding, exiting early during inference inevitably reduces accuracy compared to full-capacity embedding.
Fortunately, while coarse-grained embeddings may not achieve precise top-1 retrieval, they can filter out the most likely candidates when expanding the retrieval range to top-10 as shown in Figure~\ref{fig:design-query-harsmart}.
Thus, this challenge can be alleviated by refining the coarse-grained embeddings filtered with query information.

\textit{\# Challenge 2: Unbalanced embedding distribution.}
As described in \S\ref{sec:design-predict}, different data exits at different layers, leading to unbalanced embeddings in storage.
Although each embedding is fine-tuned to approximate the full embedding, embeddings from different exit layers retain unique characteristics.
For example, samples from similar exit layers tend to have similar embedding distributions.
As a result, a query embedding from a full-capacity encoder cannot retrieve these embeddings precisely.
This phenomenon is shown in Figure~\ref{fig:design-query}.
For single-modality retrieval on the HARSMART dataset, using the full-capacity MEM to retrieve filtered embeddings results in a top-1 accuracy of only 24.9\%, 56.6\% lower than using a 2-layer query embedding, since over 99\% of samples exit before 3 layers during local embedding.
The same phenomenon occurs in the cross-modal TWITTER dataset.

\input{fig-design-query-explain.tex}
\paragraph{Speculative retrieval}
Inspired by speculative decoding~\cite{leviathan2023fast}, a popular acceleration technique for language models, we propose feeding the query embedding at different granularities to achieve balanced filtering, as shown in Figure~\ref{fig:design-query-explain}.
(1) Speculative filtering: The top $k$ candidates at each query granularity are preserved for the second round of filtering.
(2) Global verifying: The second round selects the final top $k$ candidates from all granularities. If a sample ID is duplicated, the candidate with the next highest score is preserved.
(3) Fine-grained correcting: Finally, the coarse-grained embeddings are refined using the rest of the model to generate fine-grained embeddings, which are then matched with the query for more precise retrieval.

\paragraph{Intermediate results reuse}
As shown in Figure~\ref{fig:design-plora-reuse}, the coarse-grained embedding can be reused for fine-grained embedding.
However, due to the down-sampling structure of the output head, the coarse-grained embedding cannot be directly used for fine-grained embedding.
To simplify this, we store the intermediate activations before each down-sample layer.
This approach allows reusing the superficial embedding to reduce the cost of data-aware coarse-grained embedding, improving embedding throughput by up to 30\%.
It also extends the coarse-grained embedding to fine-grained embedding without encoding from scratch, accelerating query latency by up to 70\%.

\paragraph{Cache analysis}
The drawback of this approach is the need to cache intermediate activations.
Fortunately, we can quantize them to INT4 and de-quantize them during reuse, which takes significantly less time than re-computation (around 10 ms per embedding).
During prediction, the activations can remain in RAM.
Once coarse-grained embedding begins, these cached activations replace the intermediate variables typically stored in RAM during embedding, so no additional peak memory footprint is required.
After the process ends, the activations are released sequentially.
For cache reuse in the fine-grained embedding procedure, the activations are temporarily stored in storage, which is less constrained than RAM, until the query occurs.
The loading time is approximately 1 ms for 10 activations.
Once an image is queried, it is updated to the fine-grained embedding, and its storage can be freed.

%% file: fig-design-query.tex
\begin{figure}[t]
    \centering
    \begin{minipage}[b]{0.24\textwidth}
        \centering
        \includegraphics[width=1\textwidth]{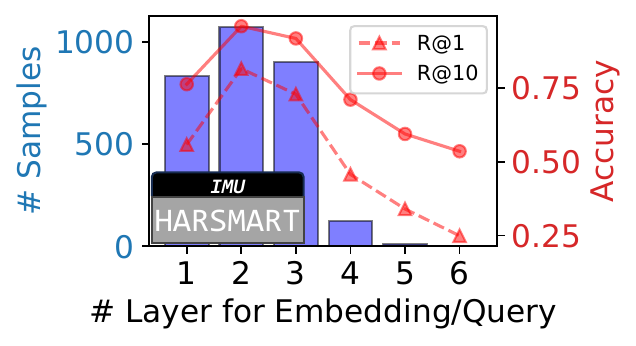}
        \vspace{-15pt}   
        \subcaption{Single modality} 
		\label{fig:design-query-harsmart}
    \end{minipage}
    ~
    \begin{minipage}[b]{0.25\textwidth}
        \centering
        \includegraphics[width=0.85\textwidth]{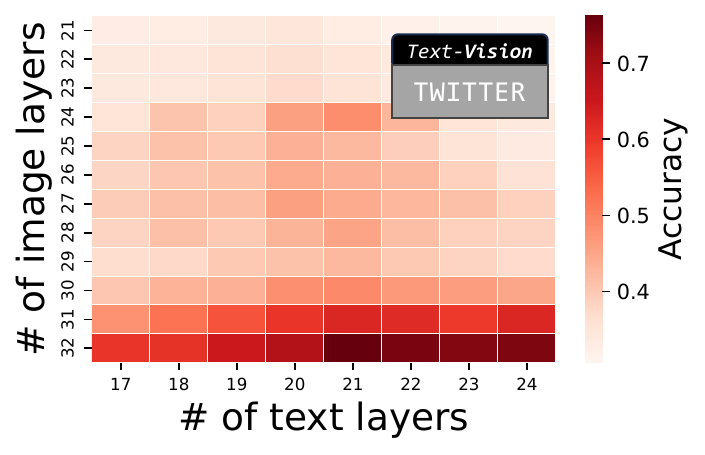}
        \vspace{-5pt}   
        \subcaption{Cross modality} 
		\label{fig:design-query-twitter}
    \end{minipage}
	\vspace{-20pt}    
    \caption{Retrieval accuracy across different embedding granularities, i.e., embeddings generated by different MEM layer exits.
    }
    \vspace{-15pt}
    \label{fig:design-query}
\end{figure}

%% file: fig-design-query-explain.tex
\begin{figure}[t]
	\centering
	 \includegraphics[width=0.48\textwidth]{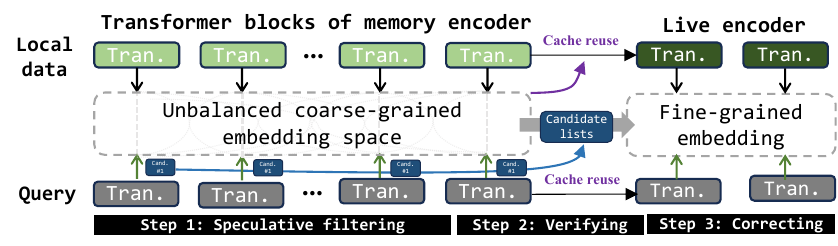}
	 \vspace{-10pt}
	 \caption{Coarse-grained embeddings are speculatively filtered. The highest-ranking embedding candidates are refined to fine-grained embeddings for final retrieval.}
	\vspace{-15pt}
	\label{fig:design-query-explain}
\end{figure}

%% file: sec-impl.tex
\section{Implementation and Setup}
\sys is built on ImageBind~\cite{girdhar2023imagebind}, an open-source multimodal embedding framework developed by Meta. 
For matching, we use matrix multiplication, as it is not the primary bottleneck in query cost. 
Further vector database optimizations, such as FAISS~\cite{qin2020efficient}, are orthogonal to \sys. 
LoRA tuning and embedding accuracy evaluations are emulated on a GPU server to enable faster iterations and energy savings. 
Embedding inference latency, power consumption, and memory usage are directly measured by running \sys on a mobile device using the open-source on-device multimodal inference engine mllm~\cite{mllm24}. 
Since mllm~\cite{mllm24} currently does not support the IMU modality and lacks GPU optimizations, we also use PyTorch on a development board for broader dataset comparisons.

\paragraph{Baselines}
We compare \sys to the following alternatives:
(1) \texttt{Multimodal Embedding Model (MEM)} without any optimization.
(2) \texttt{BranchyNet}~\cite{teerapittayanon2016branchynet}, using a traditional early-exit mechanism.
(3) \texttt{Fluid Batching}~\cite{kouris2022fluid}, a novel early-exit-aware batching algorithm that allows sample preemption at runtime.
For completeness, we also include a naive baseline without layer-wise execution, though it incurs an unaffordable memory footprint on certain mobile devices.
For a fair comparison, all baselines are equipped with ImageBind fine-tuned for the downstream task.

\paragraph{Metrics}
We evaluate the performance of \sys using the following metrics:
(1) \textit{Accuracy}: Retrieval accuracy for each task, with relative accuracy compared to the full-sized model, as shown in Table~\ref{tab:design-dataset}.
(2) \textit{Latency}: Query latency on mobile devices, defined as the time from query initiation to completion.
(3) \textit{Throughput}: The amount of content processed per second or minute, assuming all samples are buffered in storage.
(4) \textit{Energy Consumption}: Energy consumed during the embedding process.
(5) \textit{Memory Usage}: Peak memory footprint during the embedding process.

\input{tab-design-dataset.tex}
\paragraph{Dataset}
As summarized in Table~\ref{tab:design-dataset}, we use four publicly available datasets across four modalities to demonstrate the effectiveness of \sys:
(1) \texttt{COCO} dataset: Used for text-image retrieval, it contains 123k images, each paired with five captions. We use the validation subset of \texttt{COCO} to evaluate inference performance, with each caption retrieving its corresponding image.
(2) \texttt{FLICKR} dataset: Used for image-text retrieval, it consists of images paired with textual descriptions.
(3) \texttt{CLOTHO} dataset: Used for text-audio retrieval, it contains audio clips paired with textual descriptions, enabling evaluation across audio and text modalities.
(4) \texttt{HARSMART} dataset: Used for IMU retrieval, it employs fine-grained embeddings as queries to assess performance in retrieving IMU data based on embeddings.

Additionally, to demonstrate the effectiveness of \sys in real-world scenarios, we conduct a case study using recent internet data that was not seen by the model during pretraining.
Following prior empirical literature on Twitter analysis~\cite{du2020understanding}, we collect a recent publicly available dataset of Twitter memes, referred to as \texttt{TWITTER}.
The \texttt{TWITTER} dataset contains 803 images and their corresponding meme descriptions across various up-to-date topics.

\paragraph{Hardware and Quantization}
We test \sys on the NVIDIA ORIN (ORIN)~\cite{orin}, Raspberry Pi 4B (RPI4B)~\cite{rpi4b}, and a flagship smartphone with Qualcomm Snapdragon 8Gen3 (8GEN3)~\cite{rn12t}.
Since ORIN's GPU does not support INT4 computation, we load the raw model with FP32 precision.
\sys runs on the RPI4B's CPU due to the lack of CUDA support.
For the 8GEN3 smartphone, \sys runs on the CPU with the model quantized to INT4 precision to reduce memory consumption.

%% file: tab-design-dataset.tex
\begin{table}[]
    \resizebox{0.48\textwidth}{!}{%
    \begin{tabular}{|c|c|c|c|l|}
    \hline
    \textbf{Dataset}  & \textbf{Modality}   & \textbf{Size}  & \textbf{Metric} & \textbf{Perf.}           \\ \hline
    COCO~\cite{lin2014microsoft}     & Text-\textbf{Image} & 123,287    &   R@1                                                              & 0.54          \\ \hline
    FLICKR~\cite{flickr8k} & Text-\textbf{Image} & 8,091  & R@1                                                               & 0.70          \\ \hline
    CLOTHO~\cite{drossos2020clotho}   & Text-\textbf{Audio} &  3,938     &     R@10                                                            & 0.30          \\ \hline
    HARSMART~\cite{sonawane2024humanactivityrecognitionusing}  & \textbf{IMU}    & 10,299   &       Acc.                                                          & 0.78 \\ \hline
    \end{tabular}
    }
    \caption{Description of the datasets used. 
    The embedded modality is highlighted.}
    \vspace{-30pt}   
    \label{tab:design-dataset}

    \end{table}

%% file: sec-eval.tex
\section{Evaluation}
\label{sec:eval}



We evaluate \sys to address the following key questions:
(1) How much improvement does \sys achieve in terms of embedding throughput and relative retrieval accuracy under different memory budgets across various devices?
(2) How much performance improvement does each component contribute?
(3) What is \sys's performance under different query latency budgets?
(4) What is the system cost of \sys, including embedding energy and memory consumption?
(5) How does \sys perform on commodity mobile phones in daily usage scenarios?

\input{sec-eval-e2e.tex}

\input{sec-eval-ablation.tex}

\input{sec-eval-query.tex}

\input{sec-eval-cost.tex}

\input{sec-eval-case.tex}


%% file: sec-eval-e2e.tex
\subsection{End-to-end Performance}
\input{fig-eval-e2e-perf.tex}
\input{tab-eval-e2e-new.tex}

First, we present the end-to-end embedding throughput performance under the layer-wise inference setting, a more user-friendly approach for always-on daily applications due to its low memory footprint.

\textbf{\sys achieves an order of magnitude improvement in throughput.}
Table~\ref{tab:eval-e2e-perf} summarizes the embedding throughput comparison, while Figure~\ref{fig:eval-e2e-perf} shows that \sys can achieve a 14.9$\times$ average throughput improvement compared to \texttt{MEM}.
This gain is primarily driven by the early-exit mechanism, which allows the model to exit early when the embedding is sufficiently accurate, avoiding unnecessary computations.
Additionally, after parameter-efficient healing, the coarse-grained embeddings can convey similar semantics to fine-grained embeddings.
For instance, in the text-image retrieval task on the \texttt{COCO} dataset, \sys delivers an 11.7$\times$ throughput improvement with less than 3\% accuracy loss.

Regarding stronger baselines, \texttt{Fluid Batch} introduces a early-exit-aware batching mechanism, achieving a 3$\times$ throughput improvement over the naive early-exit baseline \texttt{BranchyNet} and 5$\times$ over \texttt{MEM} under the layer-wise inference setting.
However, \sys still outperforms \texttt{Fluid Batch} across all datasets, providing up to a 3$\times$ speedup in throughput.
The advantages of \sys arise not only from the early-exit mechanism but also from the pre-exit strategy, which predictively adjusts the embedding granularity based on the sample's characteristics.

Although \sys primarily targets layer-wise scenarios, we also evaluated throughput performance when loading all encoders simultaneously.
While this approach can provide significant throughput gains, it presents challenges such as out-of-memory errors on ORIN and RPI, especially for larger models like vision encoders. 
\sys maintains high throughput in a layer-wise setting, making it a more practical solution for resource-constrained devices.
For instance, on the 8GEN3 mobile, \sys can process data up to 2.5$\times$ faster than the naive MEM without loading layers sequentially, while reducing memory footprint by up to 3.3$\times$.

Interestingly, we find that healing the exited larger MEMs is more effective than using a smaller-sized foundation model. 
For example, using CLIP-b/16 with 85.6M parameters results in embeddings that are 2.7$\times$ faster than ImageBind but significantly reduces the ability to embed different modalities concisely, leading to up to 39.8\% accuracy loss.

%% file: fig-eval-e2e-perf.tex
\begin{figure*}[t]
    \centering
    \begin{minipage}[b]{1\textwidth}
        \centering
        \begin{minipage}[b]{0.24\textwidth}
            \includegraphics[width=0.97\textwidth]{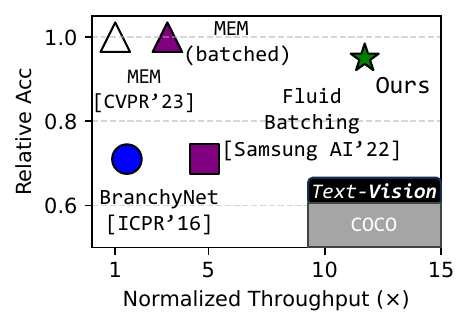}
        \end{minipage}
        ~
        \begin{minipage}[b]{0.24\textwidth}
            \includegraphics[width=0.98\textwidth]{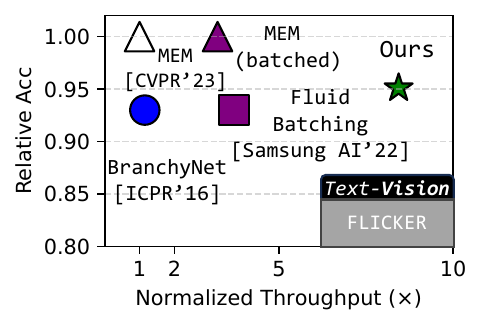}
        \end{minipage}
        ~
        \begin{minipage}[b]{0.24\textwidth}
            \includegraphics[width=0.95\textwidth]{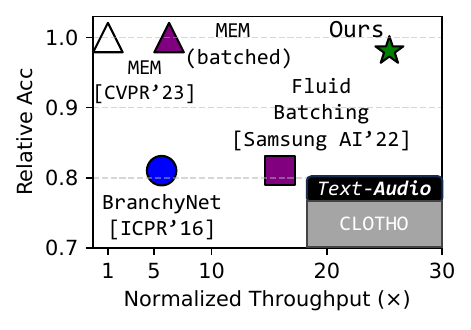}
        \end{minipage}
        ~
        \begin{minipage}[b]{0.24\textwidth}
            \includegraphics[width=0.95\textwidth]{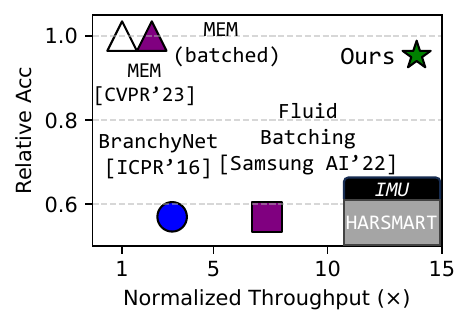}
        \end{minipage}
    \end{minipage}
    
    \vspace{-15pt}
    \caption{Illustration of throughput-to-accuracy on Jetson TX2. 
    For fairness, only layerwise baselines are included.}
    \label{fig:eval-e2e-perf}
    
\end{figure*}

%% file: tab-eval-e2e-new.tex
\begin{table*}[]
    \resizebox{\textwidth}{!}{%
    \begin{tabular}{c|cccc|cccc|cccc|cccc}
    \hline
    \textbf{Dataset} & \multicolumn{4}{c|}{\textbf{COCO}} & \multicolumn{4}{c|}{\textbf{FLICKER}} & \multicolumn{4}{c|}{\textbf{CLOTHO}} & \multicolumn{4}{c}{\textbf{SENSOR}} \\ \hline
    \multicolumn{1}{c|}{\textbf{\begin{tabular}[c]{@{}c@{}}Throughput\\ (Contents / s)\end{tabular}}} & \multicolumn{1}{c|}{\textbf{\begin{tabular}[c]{@{}c@{}}Relative\\ Accuracy\end{tabular}}} & \textbf{\begin{tabular}[c]{@{}c@{}}ORIN\\ (FP32)\end{tabular}} & \textbf{\begin{tabular}[c]{@{}c@{}}RPI4B\\ (FP32)\end{tabular}} & \textbf{\begin{tabular}[c]{@{}c@{}}8GEN3\\ (INT4)\end{tabular}} & \multicolumn{1}{c|}{\textbf{\begin{tabular}[c]{@{}c@{}}Relative\\ Accuracy\end{tabular}}} & \textbf{\begin{tabular}[c]{@{}c@{}}ORIN\\ (FP32)\end{tabular}} & \textbf{\begin{tabular}[c]{@{}c@{}}RPI4B\\ (FP32)\end{tabular}} & \textbf{\begin{tabular}[c]{@{}c@{}}8GEN3\\ (INT4)\end{tabular}} & \multicolumn{1}{c|}{\textbf{\begin{tabular}[c]{@{}c@{}}Relative\\ Accuracy\end{tabular}}} & \textbf{\begin{tabular}[c]{@{}c@{}}ORIN\\ (FP32)\end{tabular}} & \textbf{\begin{tabular}[c]{@{}c@{}}RPI4B\\ (FP32)\end{tabular}} & \textbf{\begin{tabular}[c]{@{}c@{}}8GEN3\\ (INT4)\end{tabular}} & \multicolumn{1}{c|}{\textbf{\begin{tabular}[c]{@{}c@{}}Relative\\ Accuracy\end{tabular}}} & \textbf{\begin{tabular}[c]{@{}c@{}}ORIN\\ (FP32)\end{tabular}} & \textbf{\begin{tabular}[c]{@{}c@{}}RPI4B\\ (FP32)\end{tabular}} & \textbf{\begin{tabular}[c]{@{}c@{}}8GEN3\\ (INT4)\end{tabular}} \\ \hline
    \textbf{MEM (w/o layerwise)} & \multicolumn{1}{c|}{\multirow{3}{*}{100.0\%}} & OOM & OOM & 0.17 & \multicolumn{1}{c|}{\multirow{3}{*}{100.0\%}} & OOM & OOM & 0.16 & \multicolumn{1}{c|}{\multirow{3}{*}{100\%}} & 83.3 & 0.26 & 0.34 & \multicolumn{1}{c|}{\multirow{3}{*}{100\%}} & 127 & 0.88 & / \\ \cline{1-1} \cline{3-5} \cline{7-9} \cline{11-13} \cline{15-17} 
    \textbf{MEM} & \multicolumn{1}{c|}{} & 1.92 & 0.04 & 0.05 & \multicolumn{1}{c|}{} & 1.92 & 0.04 & 0.05 & \multicolumn{1}{c|}{} & 5.23 & 0.22 & 0.27 & \multicolumn{1}{c|}{} & 31.3 & 0.74 & / \\ \cline{1-1} \cline{3-5} \cline{7-9} \cline{11-13} \cline{15-17} 
    \textbf{MEM (batched)} & \multicolumn{1}{c|}{} & 6.22 & 0.05 & 0.10 & \multicolumn{1}{c|}{} & 6.22 & 0.05 & 0.10 & \multicolumn{1}{c|}{} & 33 & 0.25 & 0.32 & \multicolumn{1}{c|}{} & 72 & 0.84 & / \\ \hline
    \textbf{BranchyNet (w/o layerwise)} & \multicolumn{1}{c|}{\multirow{3}{*}{71.0\%}} & OOM & OOM & 0.25 & \multicolumn{1}{c|}{\multirow{3}{*}{92.7\%}} & OOM & OOM & 0.19 & \multicolumn{1}{c|}{\multirow{3}{*}{81\%}} & 211 & 0.66 & 0.85 & \multicolumn{1}{c|}{\multirow{3}{*}{57\%}} & 405 & 2.81 & / \\ \cline{1-1} \cline{3-5} \cline{7-9} \cline{11-13} \cline{15-17} 
    \textbf{BranchyNet} & \multicolumn{1}{c|}{} & 2.88 & 0.06 & 0.07 & \multicolumn{1}{c|}{} & 2.21 & 0.04 & 0.06 & \multicolumn{1}{c|}{} & 29.6 & 0.55 & 0.69 & \multicolumn{1}{c|}{} & 99.9 & 2.36 & / \\ \cline{1-1} \cline{3-5} \cline{7-9} \cline{11-13} \cline{15-17} 
    \textbf{Fluid Batch} & \multicolumn{1}{c|}{} & 9.29 & 0.07 & 0.16 & \multicolumn{1}{c|}{} & 7.13 & 0.05 & 0.12 & \multicolumn{1}{c|}{} & 83.4 & 0.63 & 0.80 & \multicolumn{1}{c|}{} & 230 & 2.68 & / \\ \hline
    \textbf{Ours} & \multicolumn{1}{c|}{\multirow{2}{*}{\textbf{95.0\%}}} & \textbf{22.5} & \textbf{0.10} & 0.31 & \multicolumn{1}{c|}{\multirow{2}{*}{\textbf{95.1\%}}} & \textbf{16.2} & \textbf{0.07} & 0.22 & \multicolumn{1}{c|}{\multirow{2}{*}{\textbf{98.1\%}}} & \textbf{133} & \textbf{0.66} & 0.84 & \multicolumn{1}{c|}{\multirow{2}{*}{\textbf{95.4\%}}} & \textbf{435} & \textbf{4.52} & / \\ \cline{1-1} \cline{3-5} \cline{7-9} \cline{11-13} \cline{15-17} 
    \textbf{Ours (w/o layerwise)} & \multicolumn{1}{c|}{} & \textbf{47.9} & \textbf{0.10} & 0.33 & \multicolumn{1}{c|}{} & \textbf{33.5} & \textbf{0.07} & 0.23 & \multicolumn{1}{c|}{} & \textbf{211} & \textbf{0.66} & 0.85 & \multicolumn{1}{c|}{} & \textbf{680} & \textbf{4.71} & / \\ \hline
    \end{tabular}%
    }
    \caption{Throughput vs. relative retrieval accuracy. 
    `/' means not supported. `OOM' means out of device memory.}
    \vspace{-20pt}   
    \label{tab:eval-e2e-perf}
    \end{table*}

%% file: sec-eval-ablation.tex
\subsection{Significance of Key Designs}
\label{sec:eval-ablation}

\input{fig-eval-ablation.tex}

\paragraph{Effect of Exit Healing}
As illustrated in Figure~\ref{fig:eval-ablation}, while the zero-shot embedding of ImageBind has the generalization ability across different datasets, the exit healing mechanism is crucial for enhancing \sys's performance.
As shown by the green dotted lines, retrieval accuracy significantly improves after healing the exited branches.
For instance, compared to zero-shot \texttt{MEM}, exit healing boosts retrieval accuracy by 37.8\% and 13.2\% on average for the \texttt{COCO} and \texttt{FLICKR} datasets, respectively.

\paragraph{Effect of Data-aware Pre-exit}
After healing, \sys leverages the pre-exit mechanism to dynamically adjust embedding granularity based on each sample's characteristics. 
It can predictively exit at the optimal layer to balance the trade-off between accuracy and throughput. 
As shown in Figure~\ref{fig:eval-ablation}, compared to exiting all samples at a fixed layer, the data-aware pre-exit mechanism improves retrieval accuracy by up to 19.8\%.
The higher coarse-grained retrieval performance is crucial for achieving optimal fine-grained retrieval.

\paragraph{Effect of Speculative Fine-grained Query}
With a default query candidate pool size of 10, retrieval accuracy using filtered fine-grained embeddings is, on average, 35.5\% higher than the previous coarse-grained retrieval accuracy.
This improvement is due to the fact that over 95\% of the targets retrievable by full-sized MEMs are successfully retrieved from the toplist of coarse-grained embeddings.
As a result, the embedding accuracy of \sys is comparable to that of the full-sized MEM.


%% file: fig-eval-ablation.tex
\begin{figure}[t]
	\centering
	 \includegraphics[width=0.48\textwidth]{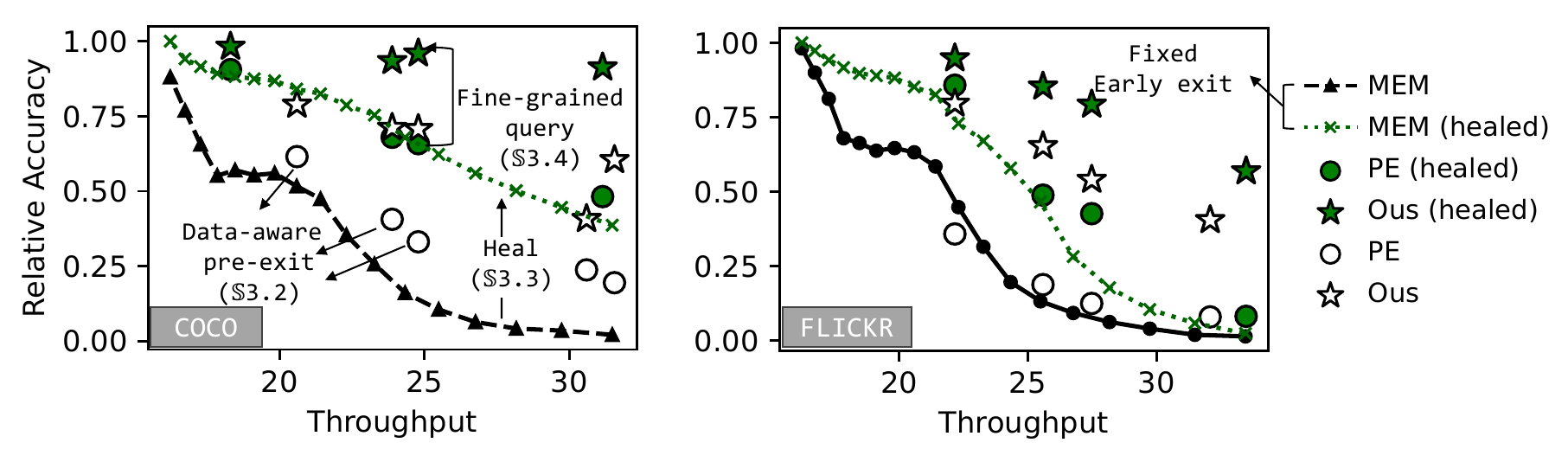}
	 \vspace{-25pt}
	 \caption{Throughput-to-accuracy trade-off with and without \sys's key designs, demonstrating their significance. 
	 PE refers to pre-exited coarse-grained embeddings without fine-grained upgrading during the query phase.}
	\vspace{-10pt}
	\label{fig:eval-ablation}
\end{figure}

%% file: sec-eval-query.tex
\subsection{Impact of Query Latency Budget}
\label{sec:eval-query}
\input{fig-eval-query.tex}

Although query cost is negligible compared to embedding cost in the long term—since queries occur less frequently than daily always-on embeddings—it is immediately noticeable to the user.
Thus, we show \sys's performance under different query latency budgets in Figure~\ref{fig:eval-query}.
Query latency consists of three components: text embedding, matching, and fine-grained embedding. 
Baseline methods with memory encoders require only the first two steps, typically taking 2 seconds.
With a higher query latency budget, we can improve fine-grained embedding accuracy from 27\% to 55\%.

Additionally, similar to web cookies~\cite{cahn2016empirical}, the query process can skip the complex fine-grained embedding when it is repeated, making it more efficient for multi-query scenarios where frequently queried items are retrieved faster.
Once a local embedding is queried, its embedding is permanently upgraded.

%% file: fig-eval-query.tex
\begin{figure}[t]
	\centering
	 \includegraphics[width=0.48\textwidth]{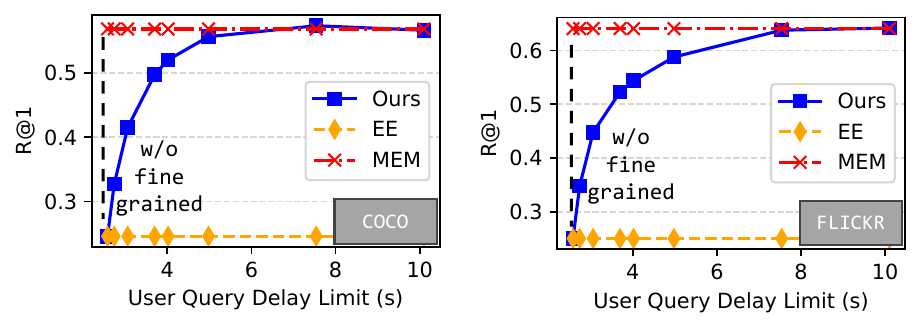}
	 \vspace{-25pt}
	 \caption{Performance under different query latency budgets.}
	\vspace{-10pt}
	\label{fig:eval-query}
\end{figure}

%% file: sec-eval-cost.tex
\subsection{System Cost}
\input{fig-eval-cost.tex}

\paragraph{Energy Consumption}
Figure~\ref{fig:eval-cost-energy} shows the normalized energy consumption of \sys and various baselines.
\sys reduces energy consumption by up to 18.2$\times$ and 13.1$\times$ on average compared to layerwise-executed baselines.
Even compared to naive MEM without layerwise execution, \sys still achieves up to 3.3$\times$ energy savings on average.
This is due to \sys's ability to determine the optimal number of layers for embedding and offload embedding computation to the less frequent querying process.

\paragraph{Memory Footprint}
The layer-wise method is much more memory-efficient than holding the entire model in memory.
This is because model weights are the primary contributor to memory footprint.
\sys is inherently memory-efficient, as it only loads the necessary layers one by one for each sample.
Compared to naive \texttt{MEM}\footnote{Memory footprint of naive \texttt{MEM} is tested in a memory-unlimited server environment.}, \sys can reduce memory usage by up to 7.7$\times$.

\paragraph{Storage Cost}
We store the embeddings of the items in INT4 precision. 
Each embedding is 1024-dimensional, resulting in a storage cost of approximately 5KB per item. 
Based on the trace statistics in \S\ref{sec:bkgnd-sys-model}, typical users encounter around 6000 images daily. 
Thus, the storage cost for image embeddings is roughly 29.3MB per day. 
Annually, this amounts to about 10.4GB, which is comparable to the storage required for a high-quality movie. 
In contrast, the current off-the-shelf solution Rewind~\cite{rewind} consumes 14GB of storage per month on average, as officially reported~\cite{rewind-compression}.

%% file: fig-eval-cost.tex
\begin{figure}[t]
    \centering
    \begin{minipage}[b]{0.48\textwidth}
        \centering
        \includegraphics[width=0.9\textwidth]{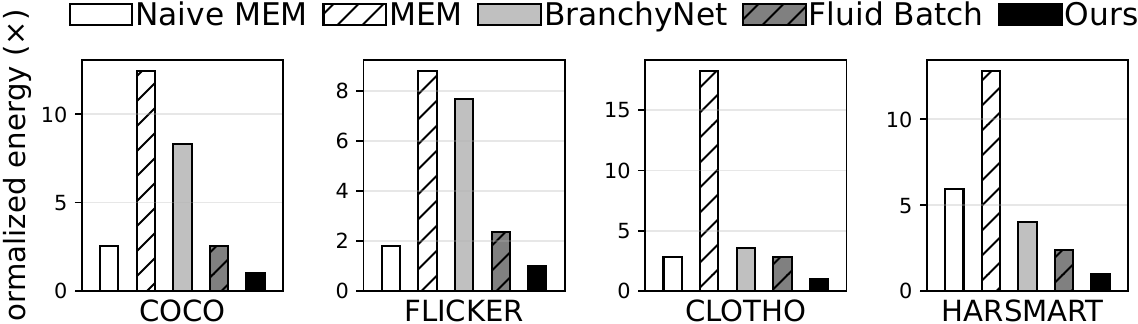}
        \vspace{-5pt}
        \subcaption{Energy consumption.} 
		\label{fig:eval-cost-energy}
    \end{minipage}
    
    \begin{minipage}[b]{0.48\textwidth}
        \centering
        \includegraphics[width=0.9\textwidth]{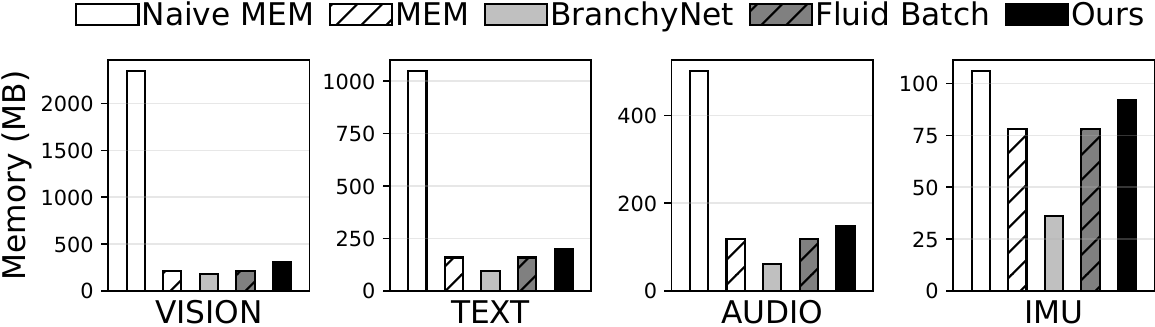}
        \vspace{-5pt}
        \subcaption{Memory footprint.} 
		\label{fig:eval-cost-memory}
    \end{minipage}
	\vspace{-25pt}    
    \caption{System cost.}
    \vspace{-10pt}
    \label{fig:eval-cost}
\end{figure}

%% file: sec-eval-case.tex
\subsection{Case Study: Twitter Meme Retrieval}
\label{sec:eval-case}

\input{fig-eval-case-result.tex}
As with the previous datasets, we evaluated the performance of \sys on the \texttt{TWITTER} dataset. 
A total of 828 figures were embedded according to the trace data collected. 
Naive \texttt{MEM} takes over an hour to complete the retrieval task on a fully utilized CPU. 
In comparison, \sys achieves a 4$\times$ throughput improvement, completing the task within 27 minutes. 
Moreover, \sys demonstrates significant resource savings, using 5$\times$ less memory and 10$\times$ less energy than the baseline.
This is due to sequentially loading layers and reducing the total number of layers executed. 
Storing these figure embeddings requires approximately 3MB, which is comparable to the size of a single raw image.
During the query phase, the query latency is 0.5s, which is acceptable for daily use, as surveyed.
Our case study demonstrates that \sys can provide high-quality embedding representation with highly optimized system performance in real-world usage scenarios.

\subsection{User Study: Mobile Application Trace}

\input{fig-eval-trace.tex}


To further validate \sys, we conducted a user study by collecting real user data and simulating the system's performance in embedding images generated during daily mobile app usage\footnote{We do not account for charging time or the energy used by the applications themselves to provide a more straightforward comparison between naive MEM and \sys.}.
Without \sys, the naive MEM system would require more than 6 battery charges per day, and over 20\% of the images would remain unembedded due to time constraints.
In contrast, \sys reduces the number of required charges by 3$\times$, allowing all daily generated data to be embedded.
This user study highlights \sys's ability to efficiently manage and embed large volumes of data, reducing the burden on battery life and ensuring that the vast majority of daily usage data is preserved and embedded in real-time.

%% file: fig-eval-case-result.tex
\begin{figure}[t]
	\centering
	 \includegraphics[width=0.43\textwidth]{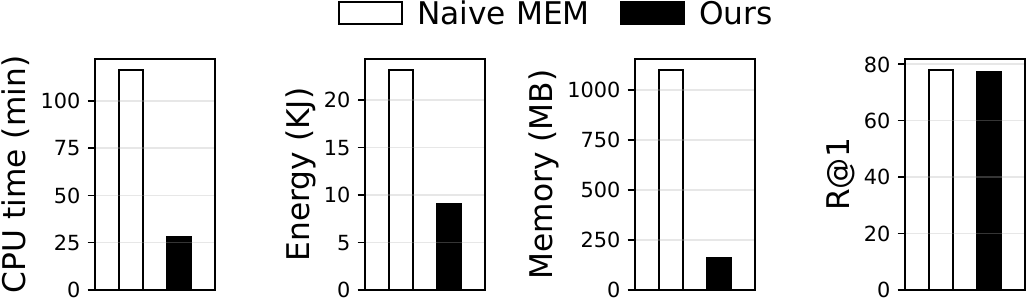}
	 \vspace{-10pt}
	 \caption{Performance analysis during 30 minutes of Twitter browsing. 
	 Device: 8GEN3~\cite{rn12t}.}
	\vspace{-10pt}
	\label{fig:eval-case-result}
\end{figure}

%% file: fig-eval-trace.tex
\begin{figure}[t]
    \centering
    \begin{minipage}[b]{0.48\textwidth}
        \centering
        \includegraphics[width=1\textwidth]{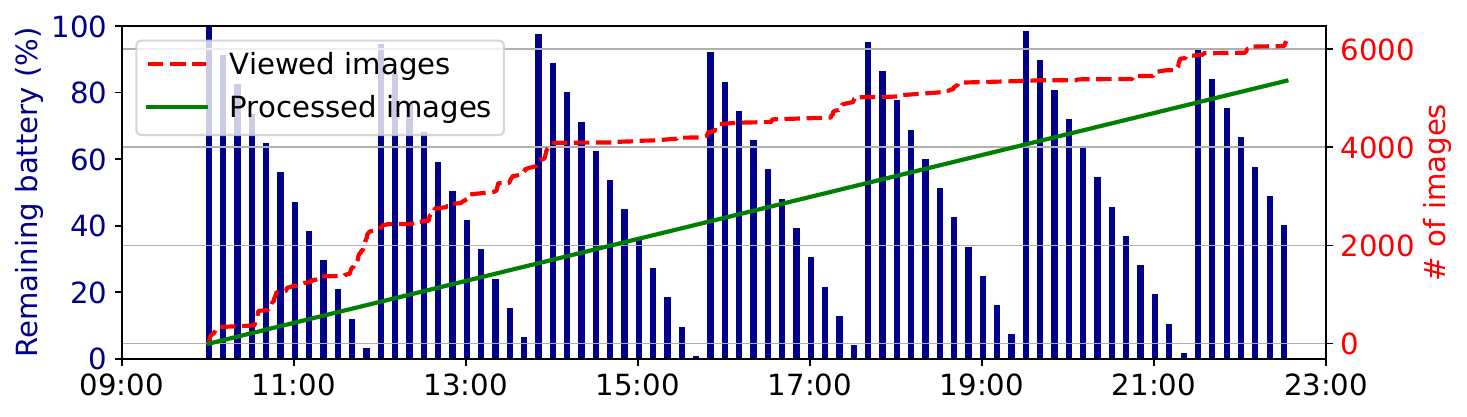}
        \vspace{-15pt}   
        \subcaption{Naive MEM.} 
		\label{fig:eval-trace-baseline}
        \vspace{0pt}
    \end{minipage}

    \begin{minipage}[b]{0.49\textwidth}
        \centering
        \includegraphics[width=1\textwidth]{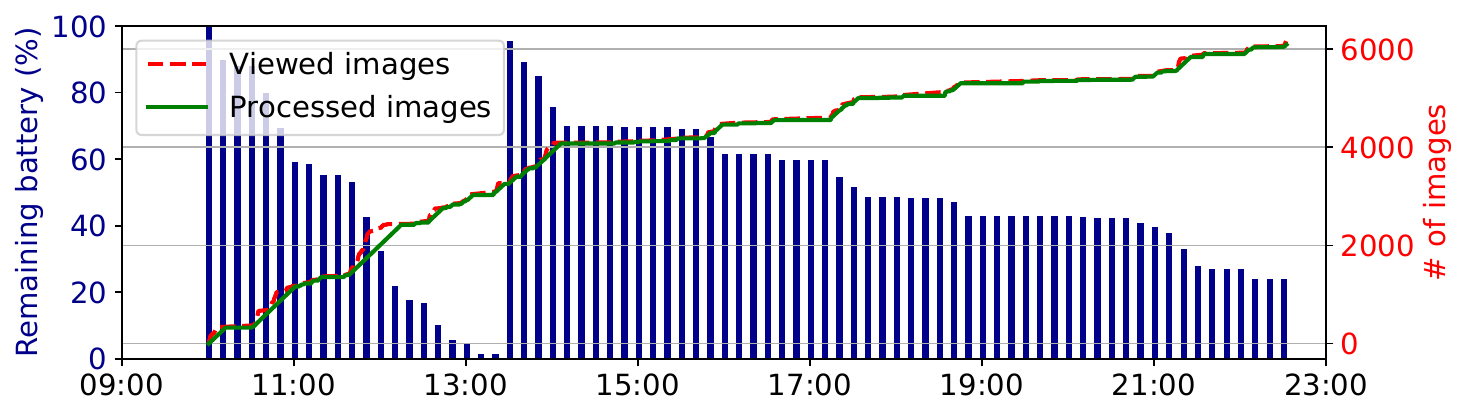}
        \vspace{-15pt}   
        \subcaption{Ours} 
		\label{fig:eval-trace-ours}
    \end{minipage}
	\vspace{-25pt}    
    \caption{Energy and throughput comparison of embedding images viewed under real mobile traces.}
    \vspace{-15pt}
    \label{fig:eval-trace}
\end{figure}

%% file: sec-related.tex
\section{Related Work}
\label{sec:related}

\paragraph{Early-Exiting}
While early exiting has been a known technique in both traditional CNNs and recent language models~\cite{fei2022deecap,li2023predictive,teerapittayanon2016branchynet,gromov2024unreasonable}, it is rarely integrated with MEMs for mobile devices.
BranchyNet~\cite{teerapittayanon2016branchynet} showed that features learned in early layers are often sufficient for classification, with only a few difficult samples requiring deeper layers. 
DynExits~\cite{wang2019dynexit} introduced learnable early-exit branch weights to avoid the pitfalls of manually defined loss weights, while DVABatch~\cite{cui2022dvabatch} dynamically adjusted batch sizes to allow independent query exits, though with limited improvement.
Layer Skip~\cite{elhoushi2024layer} and DeeCap~\cite{fei2022deecap} utilized early exits for tasks like text decoding and image captioning. 
Gromov et al.~\cite{gromov2024unreasonable} demonstrated that removing deeper layers often does not degrade performance due to the similarity between adjacent deep layers. 
In this work, we propose the first early-exiting system for on-device MEMs, providing a lightweight and efficient solution for mobile devices.

\paragraph{Predictive Early Exit}
Predictive Exit~\cite{li2023predictive} designed a low-cost prediction engine for CNNs, using zero padding, filter generation, and one-dimensional convolution to predict exit points in computer vision tasks. 
Dong et al.~\cite{dong2022resource} introduced an exit predictor that uses depthwise separable convolutions to generate scores for deciding whether to skip certain exits, reverting to confidence-based decisions when necessary. 
However, these methods are complex and not suited for attention-based transformers, where matrix multiplication dominates. 
In summary, while effective for CNNs, these approaches do not scale well to transformer-based models due to their convolution-specific designs. 
Hamed et al.~\cite{hamed2024multimodal} integrated early exits at the end of encoders to balance predictive performance and computational efficiency, but they did not address exit timing or hardware compatibility.
Our system introduces a hardware-friendly, lightweight predictor for efficient early exits in transformers, tailored for mobile devices, ensuring both performance and accuracy.

\paragraph{Multimodal Embedding Model}
According to Zhang et al.~\cite{zhang2024mm}, over 80\% (35 out of 43) of multimodal foundation models utilize ImageBind or its subset, CLIP, as their modality encoder. 
This widespread adoption highlights the efficiency and effectiveness of ImageBind in managing multimodal data.
Further optimizations have enhanced the fusion of vision and text~\cite{tang2023you,hamed2024multimodal}, as well as the fusion of dynamic sensing~\cite{hor2023sense}.
However, these methods do not address new modalities in open-set recognition, as handled by ImageBind.
For the first time, we enable efficient multimodal embedding in a unified space with usable search accuracy on mobile devices.

%% file: sec-conclusion.tex
\section{Conclusion}
\label{sec:conclusion}

We propose \sys, a novel MEM-empowered mobile service that enables user-friendly searching for multimodal mobile data. 
To improve the efficiency of on-device multimodal embedding, \sys is built on early exiting techniques to generate coarse-grained embeddings.
\sys introduces three key optimizations: predicting exits, healing exited branches, and fine-grained retrieval. These enhancements adapt traditional early-exit methods for mobile MEMs, resulting in higher embedding throughput, improved embedding quality, and better retrieval precision.
Our extensive experiments demonstrate that \sys significantly accelerates the multimodal embedding process while ensuring accurate searches.